\documentclass[twocolumn,showpacs,preprintnumbers,amsmath,amssymb]{revtex4}

\usepackage{graphicx}
\usepackage{dcolumn}
\usepackage{bm}

\begin{document}


\title{Correlations and Characterization of Emitting sources}

\author{G. Verde}
\affiliation{Istituto Nazionale di Fisica Nucleare, Sezione di Catania, Catania, Italy}
\email{verde@ct.infn.it}
\author{A. Chibihi}
\affiliation{GANIL, Caen, France}
\author{R. Ghetti}
\author{J. Helgesson}
\affiliation{School of Technology and Society, Malm{\"o} University, 
            SE-20506 Malm{\"o}, Sweden}



\begin{abstract}
Dynamical and thermal characterizations of excited nuclear systems produced during the collisions between two heavy ions at intermediate incident energies are presented by means of a review of experimental and theoretical work performed in the last two decades. Intensity interferometry, applied to both charged particles (light particles and intermediate mass fragments) and to uncharged radiation (gamma rays and neutrons) has provided relevant information about the space-time properties of nuclear reactions. The volume, lifetime, density and relative chronology of particle emission from decaying nuclear sources has been extensively explored and has provided valuable information about the dynamics of heavy-ion collisions. 
Similar correlation techniques applied to coincidences between light particles and complex fragments are also presented as a tool to determine the internal excitation energy of excited primary fragments as it appears in secondary-decay phenomena.
\end{abstract}

\pacs{Valid PACS appear here}
\maketitle

\section{INTRODUCTION}

Heavy-ion collisions are the only terrestrial means to explore the properties of nuclear matter under extreme conditions. In order to extract such nuclear matter properties, a clear understanding of the complex dynamics of heavy-ion collisions is required \cite{ref1,ref2,ref3,ref4,ref5,ref6,ref7,ref8,ref9,ref10}. The detected particles are indeed produced by different emission mechanisms and at different stages whose experimental identification
is challenging. Researchers have therefore intensively focused on obtaining a well defined dynamical and thermal characterization of particle and fragment emitting sources. 

Where and when are fragments produced? What are their thermal properties, i.e. excitation energy, internal temperature, or spin? 
At what density do nuclear multifragmentation phenomena occur? Can we learn something about their link to a liquid-gas phase transition in nuclear matter  \cite{ref11,ref12,ref13,ref14,ref15}? 

In this chapter we will present a review of those research activities that have been devoted to finding answers to these questions. We will first focus on the space-time characterization of particle emitting sources, namely the estimation of their sizes, shapes, densities, lifetimes and emission chronology. This task has been extensively addressed with intensity-interferometry studies by exploring light particle-light particle and IMF-IMF (Intermediate-Mass Fragment) correlation functions. The last section will be devoted to the thermal characterization of emitting sources by means of light particle-IMF correlation function techniques, providing information about fragment internal excitation energies and the relative proportion of the thermal component. We will finally conclude with some remarks and perspectives for future research in this field.

\section{Intensity interferometry and light particle emission}

The space-time properties of heavy-ion collisions can be accessed by intensity interferometry  \cite{ref16,ref17,ref18}. This technique was originally introduced in astronomy by Hanbury-Brown and Twiss to measure astronomical distances, such as the radii of stars and galaxies  \cite{ref19,ref20}. The technique was later extended to subatomic physics by Goldhaber et al. who studied distributions of $K$ mesons in proton-antiproton annihilation processes  \cite{ref21}. Due to their bosonic nature, two-pion correlation functions show an enhancement at zero-relative momentum. The width of this enhancement provided information about the volume of the region emitting pions in the studied processes. Pion-pion interferometry plays still today a key role in the study of heavy-ion collisions at ultra-relativistic energies where it is an important observable to investigate the production of the Quark Gluon Plasma \cite{ref22}. 

The use of intensity interferometry in heavy-ion collisions at intermediate energies is generally characterized by a more complicated scenario, as compared to the case of astronomical applications. During a nuclear reaction not only photons but several particle species can be emitted: neutrons, protons, complex particles or fragments. These emitted radiations can be either bosons or fermions, interacting with one another by means of repulsive Coulomb and attractive nuclear forces. Another important complication inherent to heavy-ion collisions is represented by the fact that the produced nuclear systems live for a very short time ranging between $10^{-22}$ and $10^{-15}$ seconds. This situation is very different from the case of astronomical objects, where the geometry of a static object is studied. Moreover, in nuclear reactions different particles can be produced at different times and by different sources. Therefore, only a full space-time characterization of all these multiple emitting sources can improve our understanding of heavy-ion collision dynamics.

\subsection{Measuring two-particle correlation functions}

\begin{figure}
\centering
\resizebox{0.9\columnwidth}{!}{
  \includegraphics{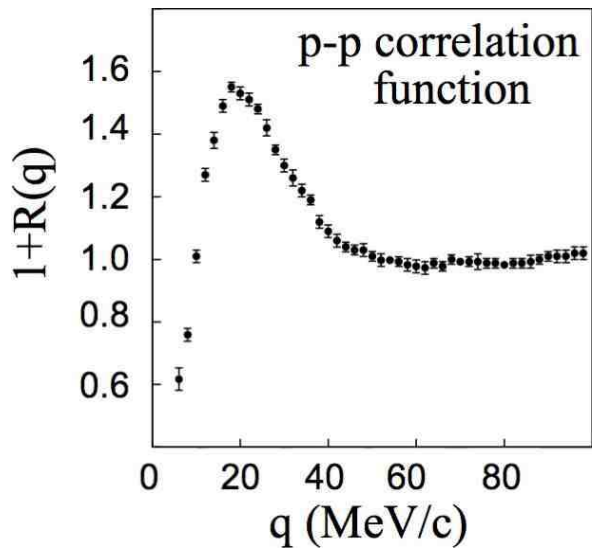}}
  \caption{Two-proton correlation function measured 
  in $^{14}$N+$^{197}$Au collisions at E/A=75 MeV (from Ref. \cite{ref27,ref28}).}
\end{figure}

\begin{figure}
\centering
\resizebox{0.85\columnwidth}{!}{
  \includegraphics{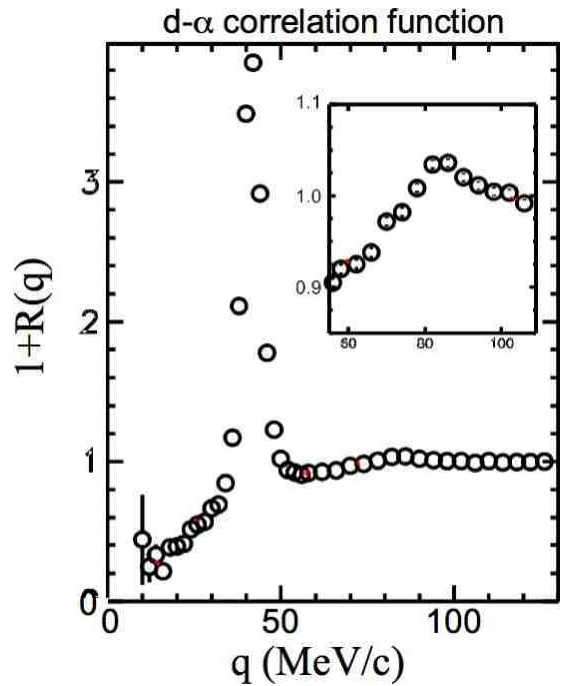}}
  \caption{Deuteron-alpha correlation function measured in Sn+Sn collisions at E/A=50 MeV with the LASSA detector array  \cite{ref29}.}
\end{figure}

Given two particles with momenta $\vec{p}_{1}$ and $\vec{p}_{2}$, total momentum $\vec{P}=\vec{p}_{1}+\vec{p}_{2}$ and momentum of relative motion $\vec{q}=\mu\left(\vec{p}_{1}/m_{1}-\vec{p}_{2}/m_{2}\right)$, the two-particle correlation function, $1+R\left(\vec{q},\vec{P}\right)$ is defined experimentally by the following equation:
\begin{equation}
\sum Y_{12}\left(\vec{p}_{1},\vec{p}_{2}\right)=C_{12}\cdot\left[1+R\left(\vec{q},\vec{P}\right)\right]\cdot\sum Y_{1}\left(\vec{p}_{1}\right)\cdot Y_{2}\left(\vec{p}_{2}\right)
\end{equation}
In this equation, $Y_{12}\left(\vec{p}_{1},\vec{p}_{2}\right)$  is the two-particle coincidence yield while   and $Y_{1}\left(\vec{p}_{1}\right)$ and $Y_{2}\left(\vec{p}_{2}\right)$  are the single-particle yields. The normalization constant $C_{12}$ is commonly determined by the requirement  $R\left(\vec{q}\right)=0$ at large relative momentum values, $q$. In order to obtain sufficient statistics, the sums in Eq. (1) are performed over all detector and particle energy combinations satisfying a specific gating condition. Experimental studies have therefore focused on two types of observables: {\it directionally-gated} and {\it angle-averaged} correlation functions. Directionally-gated correlation functions are constructed by selecting particle pairs with specific conditions on the relative direction between the relative momentum,  $\vec{q}$, and the total momentum, $\vec{P}$ \cite{ref23,ref24}. For instance, correlation functions with the vector $\vec{q}$ either parallel or perpendicular to the vector $\vec{P}$ have  been extensively constructed \cite{ref25,ref26}. Studies with such directional gates generally require high statistics. Directional effects might indeed be small and difficult to analyze \cite{ref25}. 
Alternatively, one can study angle-averaged correlation functions by integrating over the relative angle between the vectors $\vec{q}$ and $\vec{P}$. The resulting correlation function depends only on the magnitude of the relative momentum, $q$: 
\begin{equation}
\sum Y_{12}\left(\vec{p}_{1},\vec{p}_{2}\right)=C_{12}\cdot\left[1+R\left(q\right)\right]\cdot\sum Y_{1}\left(\vec{p}_{1}\right)\cdot Y_{2}\left(\vec{p}_{2}\right)
\end{equation}
Experimentally, the product of the single yields, $Y_{1}\left(\vec{p}_{1}\right)\cdot Y_{2}\left(\vec{p}_{2}\right)$ in Eqs. (1) and (2), has often been approximated with the uncorrelated two-particle yields, $Y_{12}^{unco}\left(\vec{p}_{1},\vec{p}_{2}\right)$, constructed via the so-called event-mixing technique: particles 1 and 2 are taken from two different events and the correlation function is calculated as $1+R(q)=C_{12}\cdot Y_{12}/Y_{12}^{unco}$ . 

Figures 1 and 2 show examples of angle-averaged 
\linebreak[4]
proton-proton and deuteron-alpha correlation functions, respectively, represented as a function of the relative momentum $q$ between the particle pairs \cite{ref29,ref30,ref31}. If the two emitted particles were totally uncorrelated, the probability of detecting them in coincidence would be equal to the product of the probabilities of detecting the singles, i.e. $Y_{12}\left(\vec{p}_{1},\vec{p}_{2}\right)\approx Y_{1}\left(\vec{p}_{1}\right)\cdot Y_{2}\left(\vec{p}_{2}\right)$, resulting in a flat correlation function, $R(q)=0$, at all $q$-values. Experimentally it is easily observed that this is not the case. As one detects pairs at small relative momentum, strong deviations from unity are observed. These deviations are due to quantum statistics and to the so-called final state interactions (FSI) \cite{ref23}. In the case of identical fermions (bosons), the relative wave function must respect anti-symmetrization (symmetrization) rules that induce measurable effects in the correlation function at small relative momenta \cite{ref19,ref20,ref21,ref23}. Furthermore, the coincident particles can interact with their mutual Coulomb and nuclear interaction. The Coulomb repulsion is responsible for the anti-correlation at small q-values. The nuclear attractive force is responsible for the observed prominent peaks both in $p-p$ and $d-\alpha$ correlation functions. 

At large relative momenta, $q\approx\infty$, the correlation functions shown in Fig. 1 and 2 appear as flat, $R(q)\approx 0$, indicating the absence of correlations between the coincident particles. However, the presence of collective motion can generate correlation effects even at large relative momenta where the correlation function may significantly deviate from the limit, $R(q)\approx 0$. If collective motion exists, the uncorrelated relative momentum distribution, constructed by mixing particles from different events, can contain additional collective components that do not exist in the coincidence spectrum. These additional contributions may affect the correlation function constructed from the ratio of the coincidence and the uncorrelated spectra \cite{ref37,ref25a}. 

In general, the shape of the correlation function is sensitive to the space-time properties of particle emitting sources produced during the reaction \cite{ref23}. In order to properly access information about these emitting sources, the use of detector arrays with a high angular and energy resolution is required. Especially the angular resolution plays an important role in determining the exact location and the shape of the resonance peaks and in accessing the correlation function at very low relative momentum. In this respect, position sensitive detectors and silicon strips have been quite successful thanks to their capability of providing relative angle measurements as small as 0.1$^o$-0.3$^o$.

\section{Space-time properties from two-proton correlation functions}

\begin{figure}
\centering
\resizebox{0.8\columnwidth}{!}{
  \includegraphics{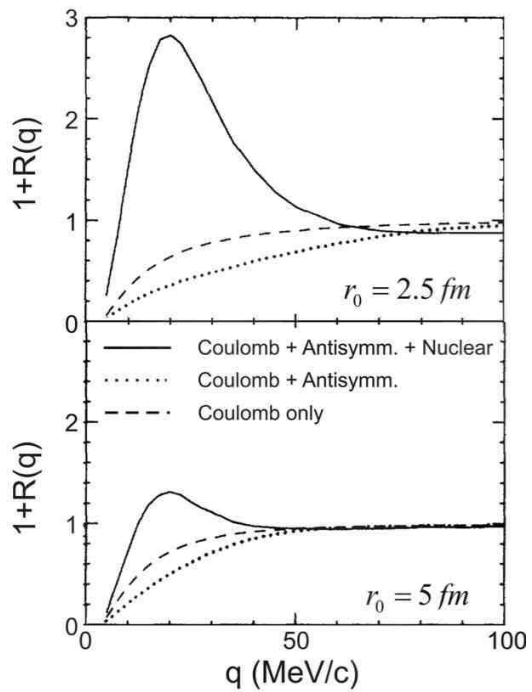}}
  \caption{Two-proton correlation functions calculated by means of Eq. (3) assuming a Gaussian spherically symmetric source function, $S\left(\vec{r}\right)\propto\exp\left(-r^{2}/r_{0}^{2}\right)$, with source size $r_{0}$=2.5 (top panel) and 5 fm (bottom panel). The different lines correspond to calculations performed by considering the anti-symmetrization and all the final state interactions (solid line), only the Coulomb interaction (dashed line), and the Coulomb interaction and the anti-symmetrization of the wave function (dotted line). }
\end{figure}

Intensity interferometry has extensively been used with protons, these particles being abundantly produced at all incident energies and easily detected with high resolution. Theoretically, the proton-proton correlation function is calculated by the so-called {\it Koonin-Pratt equation} (KP equation) \cite{ref24}: 

\begin{equation}
1+R\left(\vec{q}\right) = 1+\int d\vec{r}\; S\left(\vec{r}\right)\cdot K\left(\vec{r},\vec{q}\right)
\end{equation}

The goal of intensity interferometry consists of solving Eq. (3): from the measured correlation function on the l.h.s of Eq. (3) one needs to extract the unknown {\it source function}, $S\left(\vec{r}\right)$. This source function is defined as the probability of emitting two particles at relative distance $\vec{r}$, calculated at the time when the last of the two particles is emitted. The so-called {\it kernel function}, $K\left(\vec{r},\vec{q}\right)$, can be calculated as $K\left(\vec{r},\vec{q}\right)=\left| \Psi_{\vec{q}}\left(\vec{r}\right)\right|-1$, where $\Psi_{\vec{q}}\left(\vec{r}\right)$ is the proton-proton scattering wave function  \cite{ref19,ref20,ref24,ref30}. The kernel contains all the information about the anti-symmetrization of the proton-proton wave function, due to their Fermionic nature, and the mutual Coulomb and nuclear final state interactions (FSI). 

\begin{figure*}
\centering
\resizebox{1.6\columnwidth}{!}{
  \includegraphics{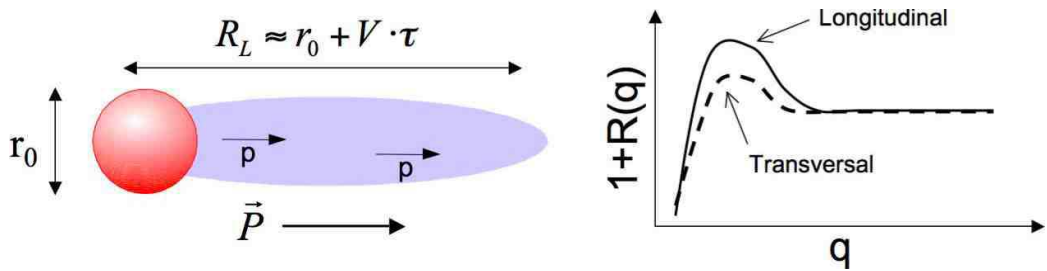}}
  \caption{Effects of finite emission lifetimes on two-proton source functions (left side) and directionally-gated correlation functions (see text for details).}
\end{figure*}

Figure 3 shows correlation functions calculated with Eq. (3) by using a 
Gaussian-shaped source function, 
\linebreak[4]
$S\left(\vec{r}\right)\propto\exp\left(-r^{2}/r_{0}^{2}\right)$, with width parameter values of $r_{0}$=2.5 fm (small source, top panel) and $r_{0}$=5 fm (large source, bottom panel) \cite{ref16}. Gaussian sources have extensively been used in the literature due to their simplicity. The dashed lines in Fig. 3 correspond to the correlation functions obtained for protons interacting only with the mutual Coulomb force. This repulsive interaction between protons is responsible for the anti-correlation at small relative momentum, 
q$<$15~MeV/c. The dotted line is obtained by adding the two-fermion anti-symmetrization in the two-proton wave function, inducing a further anti-correlation in the region q=15-60 MeV/c, due to the Pauli principle that prevents protons from occupying relative momentum states over an interval of $\Delta p_{x}=h/\Delta x$, where $\Delta x$ is the spread of the proton spatial distribution in a certain direction $x$. Finally, when also the final state mutual nuclear interaction is included, the $p-p$ correlation functions displays a prominent peak at 20 MeV/c. This peak is due to the s-partial wave of the proton-proton scattering problem and strongly depends on the volume of the emitting source. Indeed, when a large emitting source is used (bottom panel in Fig. 3), the peak  height reduces significantly. 

The correlation functions displayed in Fig. 3 have been determined assuming that the two protons are simultaneously emitted by a source with zero lifetime. However, proton emission is known to occur over finite timescales (ranging from few tens of fm/c in the case of pre-equilibrium emission to thousands of fm/c in secondary decay processes). If the particle pairs are not emitted simultaneously, the source function is affected by a space-time ambiguity and will appear deformed, as it is schematically shown in Fig. 4 (left panel) \cite{ref23,ref25}. If $r_{0}$ is the actual geometrical source size, the source function appears elongated and with a larger size in the direction defined by the total momentum vector $\vec{P}$. The elongation is approximately given by $V\tau$, with $V$ and $\tau$ being, respectively, the average pair velocity and the emission lifetime. Due to this deformation of the source function, the Pauli suppression effect, described in Fig. 3 (dotted lines), plays a key role in determining the shape of the correlation function. As it is schematically shown in the right panel of Fig. 4, the transverse correlation function, constructed with the relative momentum, $\vec{q}$, perpendicular to the total momentum, $\vec{P}$, will undergo a larger Pauli suppression as compared to the correlation function constructed by selecting a relative momentum, $\vec{q}$, parallel to the total momentum, $\vec{P}$. This simple qualitative argument shows that directionally-gated correlation functions allow one to disentangle the space and time information hidden in the source function, providing quantitative estimates of both source size and lifetime \cite{ref25,ref26}.

\subsection{Volumes and Lifetimes from Directionally-Gated correlation functions}

The sensitivity of directional correlation functions to finite lifetimes and emission volumes has stimulated an extensive research activity in the field of heavy-ion collisions. 

\begin{figure}
\centering
\resizebox{0.9\columnwidth}{!}{
  \includegraphics{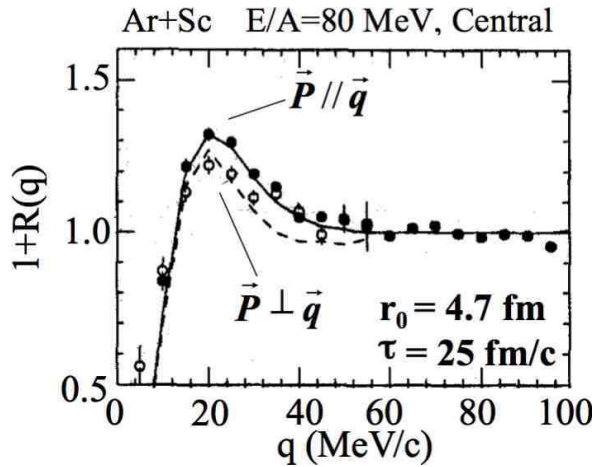}}
  \caption{Data points: directional two-proton correlation functions measured in Ar+Sc collisions at E/A=80 MeV (from Ref. \cite{ref25}). The solid and dashed lines correspond, respectively, to longitudinal and transversal calculated correlation functions.}
\end{figure}

Figure 5 shows two-proton correlation functions measured in central Ar+Sc collisions at E/A=80 MeV and for proton pairs with total momenta P=400-600 MeV/c \cite{ref25}. The filled and open circles correspond, respectively to longitudinal and transverse correlation functions. A stronger Pauli suppression in the transverse direction is observed, as compared to the longitudinal direction. These results are one of the first experimental evidences of the predicted lifetime effects in two-proton correlation functions. In the same reaction, protons at higher total momenta showed no directional dependence \cite{ref31}, consistent with negligible lifetime effects. Indeed, these fast protons are dominated by rapidly decaying pre-equilibrium sources. The directional correlations in Fig. 5 were analyzed by using the Koonin-Pratt equation, Eq. (2), assuming a source function composed by a Gaussian spatial distribution and an exponentially decaying time profile, $S\left(\vec{r},t\right)\propto\exp\left(-r^{2}/2r^{2}_{0}\right)\cdot\exp\left(-t/\tau\right)$. The simultaneous best-fit of longitudinal and transverse correlations provided source radii and lifetimes of $r_{0}$=4.5-4.8 fm and $\tau$=10-30 fm/c, respectively. The obtained geometrical size, $r_{0}$=4.5-4.8 fm, is comparable to the size of the overlapping region in central Ar+Sc collisions. The extracted short lifetime was interpreted as consistent with proton sources dominated by the early pre-equilibrium stage of the reaction. These results were also predicted by BUU transport models that are expected to well describe the dynamical early stages of the reaction  \cite{ref31}.

Directional correlation functions have also been used to characterize proton evaporation from long-lived nuclear systems at relatively low excitation energies \cite{ref32,ref33}. In the study of quasi-compound projectiles produced in $^{129}$Xe + $^{27}$Al reactions at E/A=31 MeV \cite{ref32}, a lifetime of the order of $\tau\approx$~1300 fm/c and a source size of $r_{0}\approx$~2.2 fm were obtained. However, this source size appeared small as compared to the size of the evaporating compound system. The difficulty in explaining the results was attributed by the authors to possible model dependences in the source parametrization. Similar difficulties have been encountered in the study of $^{32}$S+$^{nat}$Ag reactions at E/A = 22.3 MeV/u \cite{ref30}. Directional two-proton correlation functions were used to explore both pre-equilibrium emission (fast sources) and proton evaporation (slow source) from the produced compound nuclei. In both cases no significant difference between transverse and longitudinal correlation functions were observed. While this observation can be expected in the case of pre-equilibrium emission, the absence of any directional effects for low energy protons evaporated from compound nuclei appeared controversial. 

\begin{figure}
\centering
\resizebox{0.9\columnwidth}{!}{
  \includegraphics{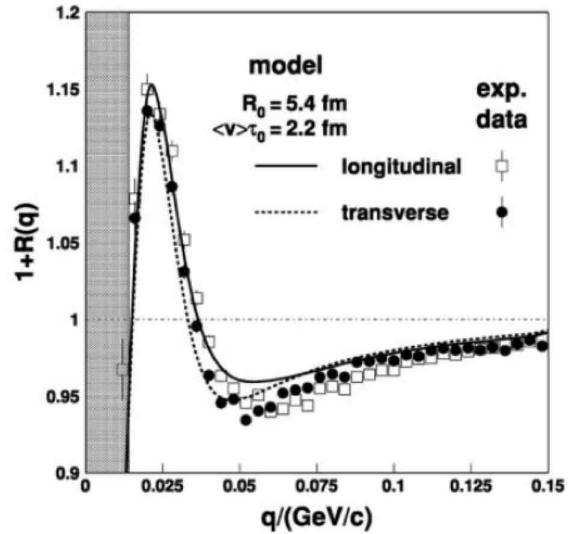}}
  \caption{Longitudinal (open dots and solid line) and transversal (close dots and dotted line) two-proton correlation functions measured in $^{96}$Zr($^{96}$Ru)+$^{96}$Zr($^{96}$Ru) central collisions at E/A=400 MeV with the FOPI detector \cite{ref26}.}
\end{figure}

\begin{figure*}
\centering
\resizebox{1.3\columnwidth}{!}{
  \includegraphics{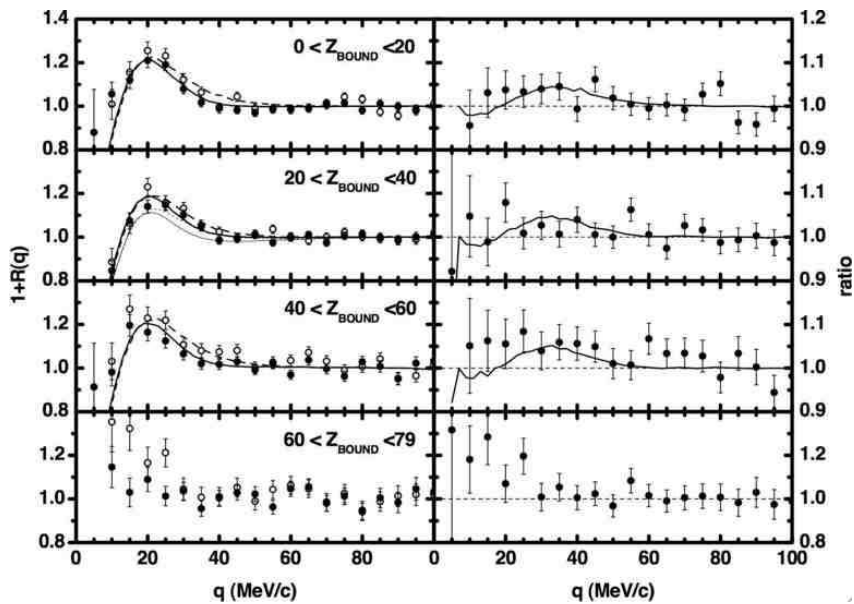}}
  \caption{Left panel: Transverse (closed symbols) and longitudinal (open symbols) two-proton correlation functions measured by the ALADiN collaboration in the decay of target spectators formed in Au+Au collisions at E/A=1 GeV \cite{ref35}. The data from the top to the bottom panel correspond to increasing impact parameter, selected by means of the Z$_{bound}$ observable \cite{ref11,ref35}. Right panel: ratio between longitudinal and transverse correlation functions.}
\end{figure*}

The space-time properties of heavy-ion collisions at relativistic incident energies were explored by the FOPI and the ALADiN collaborations \cite{ref26,ref37,ref25a,ref34,ref35,ref36} to access information about nuclear densities and to study the temporal evolution of excited nuclear systems. Figure 6 shows directional two-proton correlation functions measured in central $^{96}$Zr($^{96}$Ru)+$^{96}$Zr($^{96}$Ru) collisions at E/A=400 MeV \cite{ref26}. A very small Pauli suppression in the transverse correlation function (solid point) as compared to the longitudinal correlation function (open points) is observed. This suppression is consistent with a slightly deformed Gaussian source characterized by a negligible lifetime and a geometrical size of about $r_{0}\approx$~5.4 fm \cite{ref26}. The obtained negligible lifetime can be attributed to the presence of a strong collective motion and the consequent decrease of emission timescales in participant matter. Longer emission lifetimes can be expected in the decay of spectator matter produced in peripheral collisions \cite{ref11}. The left panels in Fig. 7 show transverse (closed symbols) and longitudinal (open symbols) two-proton correlation functions measured in the decay of target spectators produced in Au+Au collisions at E/A=1 GeV \cite{ref35}. The data from the top to the bottom are obtained by imposing different gates on the Z$_{bound}$ observable, defined as the sum of the charges of all fragments with Z$\geq$2 and emitted by projectile spectators \cite{ref11}. Z$_{bound}$ increases with increasing impact parameter and decreasing spectator excitation-energy. These directional correlation functions provided lifetimes $\tau\leq$~20 fm/c and source radii $r_{0}\approx$~8 fm, independently of Z$_{bound}$. From an estimate of the amount of nucleons in the decaying spectator it was possible to deduce densities ranging between $\rho/\rho_{0}\approx$~0.15 and $\rho/\rho_{0}\approx$~0.4. Even in the case of spectator decay, the obtained lifetimes appear to be rather short. High proton identification thresholds \cite{ref37,ref35} could result in correlation functions that are dominated by fast particles emerging from the short-lived pre-equilibrium emission stages of the reaction. On the other hand, other authors have investigated the intrinsic small sensitivity of directional correlation functions to those long lifetimes typical of evaporation processes and secondary decays. Can one probe the time structure of proton emissions when lifetimes of the order of $\tau>$10$^{3}$-10$^{4}$ fm/c exist? In Ref. \cite{ref30} it was observed that, if the emission lifetime is very long, longitudinal and transverse correlation functions become undistinguishable and, as a result, a quantitative extraction of the lifetime itself becomes difficult.

\subsection{Space-time source extents from angle-averaged two-proton correlations}

Angle-averaged correlation functions are constructed experimentally by removing any conditions on the angle between the relative and the total momentum of the emitted proton pairs. From this point of view, these observables do not require as high statistics as in the case of directionally gated correlation functions. In order to extract physics information, one needs to use the {\it angle-averaged} Koonin-Pratt equation \cite{ref17,ref23,ref27,ref28,ref38}:

\begin{equation}
R\left(q\right) = 4\pi\int dr\cdot r^{2}\; S\left(r\right)\cdot K\left(r,q\right)
\end{equation}

where $K(r,q)$ is the angle-averaged kernel. The resulting spherically symmetric source function, $S(r)$, contains information about its spatial extent and its finite lifetimes folded together in the value of the relative distance, $r$. Therefore, it is not easy to disentangle the space-time ambiguity contained in the emitting source. However, even under these limited conditions, angle-averaged correlation functions have provided space-time information about particle emission mechanisms. 

\begin{figure}
\centering
\resizebox{0.9\columnwidth}{!}{
  \includegraphics{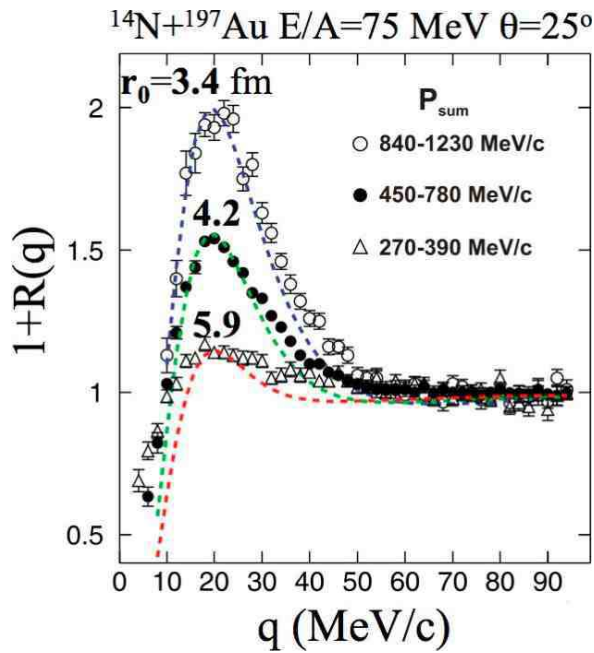}}
  \caption{Data points: Two-proton angle-averaged correlation functions measured in $^{14}$N+$^{197}$Au at E/A=75 MeV \cite{ref27,ref28}. The different symbols correspond to different gates in the total momentum of proton pairs, as indicated on the figure. The dashed lines are calculated assuming Gaussian-shaped source functions and correspond to best fits of the height of the peaks at 20 MeV/c. }
\end{figure}

In order to solve the angle-averaged Koonin-Pratt equation, Eq. (4), and extract information about the emitting source function, $S(r)$, different approaches have been proposed. In the following subsections we will present the main results obtained with {\it Gaussian source approaches} and {\it imaging techniques}.

\subsubsection{Gaussian source approaches}

In Fig. 8 angle-averaged two-proton correlation functions measured in $^{14}$N+$^{197}$Au reactions at E/A=75 MeV are shown for three different gates on the total momentum of detected proton pairs, P=270-390, 450-780 and 840-1230 MeV/c \cite{ref27,ref28}. The dashed lines correspond to best-fits of the experimental data with Eq. (4), where the source function is assumed to be characterized by a Gaussian shape, $S_{\vec{P}}\left(\vec{r}\right)\propto\exp\left(-r^{2}/2r_{0}^{2}\right)$. Gaussian source sizes, $r_{0}$=5.9, 4.2 and 3.4 were obtained, respectively, for P=270-390, 450-780 and 840-1230 MeV/c. 

\begin{figure}
\centering
\resizebox{1.0\columnwidth}{!}{
  \includegraphics{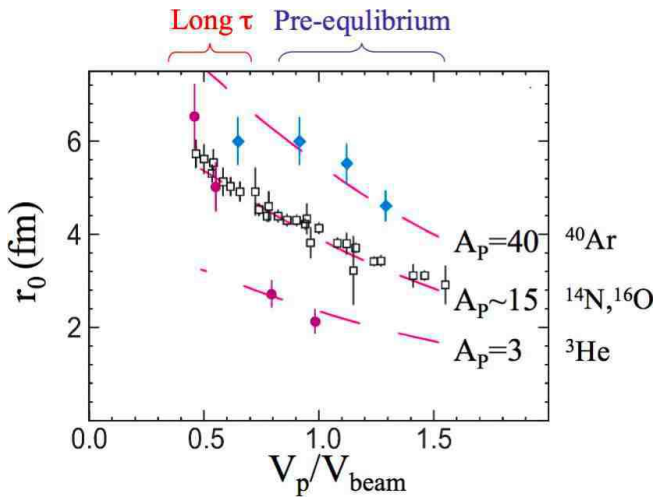}}
  \caption{ Gaussian source sizes obtained from the study of reactions induced by $^{14}$N and $^{16}$O projectiles (open symbols), $^{3}$He projectiles (closed circles) and $^{40}$Ar projectiles (closed diamonds) on $^{197}$Au targets (adapted from Refs. \cite{ref16,ref17}).}
\end{figure}

The total momentum dependence of two-proton correlation functions, shown in Fig. 8, has been extensively explored in the literature. Figure 9 shows a collection of Gaussian source sizes measured in reactions induced by different projectiles ($^{3}$He, $^{14}$N, $^{16}$O and $^{40}$Ar) impinging on Au and Ag targets \cite{ref16,ref17,ref39}. Source sizes are represented as a function of the average velocity of the coincident proton pairs, $v_{p}=1/2\left(p_{1}+p_{2}\right)/M_{p}$, normalized to the beam velocity, $v_{beam}$. The observed decrease of the size of the emitting source with increasing proton velocities has been interpreted as a consequence of the cooling dynamics of the produced nuclear systems \cite{ref16,ref17,ref28}. Energetic protons are associated with small sources as one would expect in the case of emissions at the early stages of the reaction. For less energetic protons, however, the extracted source radii increase as one would expect at the later stage when the system is cooled and expanded. The systematics of two-proton radii with the size of the interacting projectile and target nuclei is more involved. The data points corresponding to reactions induced by $^{14}$N and $^{16}$O projectiles overlap and are interpolated by a dashed line. The other two dashed lines are obtained by multiplying and dividing the $^{14}$N/$^{16}$O line by a factor proportional to $\left(A_{P}/14\right)^{1/3}$, where $A_{P}$ is the mass number of the projectile nuclei. It is clear that for high energy protons, $v_{p}/v_{beam} >$~0.5, where fast pre-equilibrium emission plays an important role, the two-proton source radius scales approximately with the radius of the projectile. This scaling with the radius of projectile nuclei is not observed in the case of low energy protons that are known to be predominantly emitted by evaporation and secondary-decay processes. The presence of such long-lifetime emissions makes two-proton correlation functions more difficult to be interpreted. This is confirmed by the difficulty in reproducing the shape of the experimental data for pairs at low total momenta (see Fig. 8). 

The Gaussian radii $r_{0}$ shown in Fig. 9 must be considered as upper bounds for the actual source sizes. Indeed, since angle-averaged source function fold space and time extents together, a small size could correspond both to a small emitting volume or to short proton emission times. On the other hands, the use of gates on total momentum or, alternatively, on proton velocities is expected to reduce the space-time ambiguities contained in the angle-averaged correlation function. Indeed, these gates should help to isolate, to some extent, more localized emitting sources. Stimulated by these ideas, the CHIC collaboration has used statistical event generators to simulate total-momentum gated correlation functions and estimate emission lifetime and spatial extents in Ni+Al reactions at E/A=45 MeV \cite{ref40}. The authors explored also the existence of a possible overlap of multiple sources with different lifetimes (corresponding to pre-equilibrium and evaporative emissions). Due to the selectivity of the total momentum gate, it was still possible to estimate both a spatial Gaussian source size, $R_{G}$=2.7$\pm$0.3, and an emission lifetime, $\tau_{p}\approx$400$\pm$200 fm/c. 

Systematic studies of source sizes in central Ca+Ca and Au+Au collisions with varying incident energies have been recently performed by the FOPI collaboration \cite{ref34}. The size of the two-proton source is observed to decrease as the incident energy is increased from E/A= 400 MeV to E/A = 1500 MeV. In the case of Au+Au reactions, the source size decreases from 5.0 fm to 4.1 fm. A smaller source size change is observed in the case of Ca+Ca collisions. This result is consistent with participant matter which is compressed to higher and higher densities as one increases the incident energy. This indication is quite attractive and further work on the details of these two-proton correlation functions can provide useful tools to investigate the nuclear equation of state. 

\begin{figure*}
\centering
\resizebox{1.5\columnwidth}{!}{
  \includegraphics{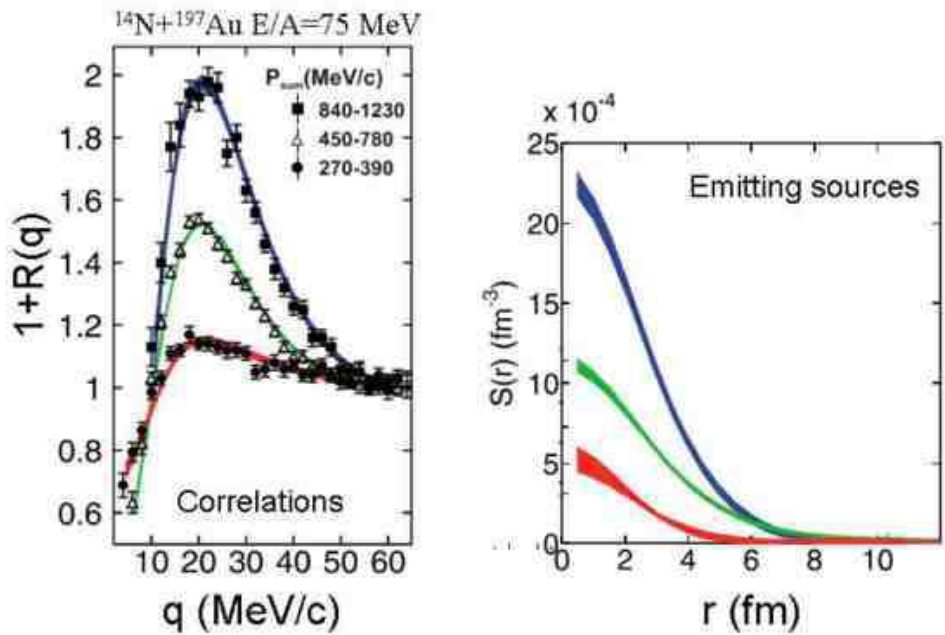}}
  \caption{Imaging analysis of data points already shown in Fig. 8 and corresponding to two-proton correlation functions measured in $^{14}$N+$^{197}$Au reactions at E/A=75 MeV \cite{ref28}. The left panel shows imaged correlation functions by thick bands. The right panel displays the extracted source function profiles \cite{ref41}.}
\end{figure*}

\subsubsection{Imaging analyses}

The importance of studying the detailed shape of two-proton correlation functions was recently addressed by introducing an imaging approach to intensity interferometry \cite{ref41,ref42,ref43,ref44}. This imaging technique consists of extracting the source profile, $S(r)$, by a numerical inversion of the Koonin-Pratt equation, Eq. (6). No a-priori assumptions about the source shape are made, thus considerably reducing model dependences. Figure 10 shows an application of the imaging technique to two-proton correlation functions measured in $^{14}$N+$^{197}$Au collisions at E/A=75 MeV \cite{ref28,ref41}. The thick lines on the left panel of the figure represent the imaged measured correlation functions. The imaging technique reproduces in details the entire shape of the correlation functions. On the right panel of Fig. 10 the extracted imaging profiles are also shown. Contrary to the case of Gaussian source analyses shown in Fig. 8, the source size does not decrease with increasing proton pair total momentum, remaining constant at about 3 fm. This result is very different from the systematics shown in Fig. 9, where the source size was mostly extracted from the height of the peak at 20 MeV/c. This constancy of the source size with increasing total momentum of the protons shows that the height of the peak at 20 MeV/c does not provide unambiguous information about the size of the source \cite{ref41}. These results have been explained within a simple scenario where protons are emitted by the overlap of a fast source, corresponding to pre-equilibrium emissions, and a slow source representing the last stages dominated by secondary decays of excited fragments \cite{ref41,ref45}. In this schematic model, fast emissions provide a fraction f of the total proton yields, $Y_{fast}=f\cdot Y$, while the slow secondary decay sources provide the remainder, $Y_{slow}=\left(1-f\right)\cdot Y$. In this limit, the profiles extracted from the imaging analysis of the experimental $R(q)$ at q$>$10 MeV/c and shown on the right panel of Fig. 10 are strongly dominated by the dynamical source. The size of this source is determined by the width or, even better, by the shape of the peak at 20 MeV/c. The detailed study of the shape provides also a measure of the relative contributions of the fast source, $f$, and of the slow source, $1-f$. This fast/slow contributions represent new physics information contained in proton-proton correlation functions, opening the opportunity to constrain the importance of secondary decays. The detailed shape of the long-lived portion of the emitting source contributes to the source function $S(r)$ at large relative distance values, $r>$10 fm, where the kernel $K(q,r)$ is dominated by the Coulomb interaction. These large distances influence the correlation function only at very small relative momenta, $q<$10 MeV/c, where measurements are difficult because of the finite angular resolution typical of the used experimental setups \cite{ref41}. Therefore, the details of these slow emitting sources $r>$10 fm  are hard to access. In Fig.~10, the observed quasi-Gaussian shapes refer only to the fast source. This is demonstrated by the fact that their integral over relative distances r is less than unity. If the slowly emitting sources could be experimentally accessed by measuring very low relative momentum particle pairs, it would be possible to image even the tails of the source profiles at large relative distances. In that case, one would clearly observe how the emitting source can deviate significantly from the simplified Gaussian shapes assumed in most works on two-proton correlation functions. Such detailed studies on the shape of two-proton correlation functions require high resolution experiments that represent a challenge for the future. 

With the introduction of imaging analyses, the systematics of proton-proton source sizes shown in Fig. 9 is expected to change. On the other hand, such high order shape analyses require high resolution measurements that certainly represent a challenging perspective for the future.

\section{Space time characterization from uncharged radiations}  

Considerable experimental efforts have also been devoted to the measurement of correlation functions involving non-charged radiations (neutrons and photons). 

Figure 11 shows two-neutron correlation functions measured in Ni+Al, Ni, Au reactions at E/A=45 MeV \cite{ref40,ref45,ref46}. These measurements are very complicated because of cross-talk problems: the interaction and registration of the same neutron in two different detectors, can substantially affect neutron coincidence measurements \cite{ref47,ref48}. 

\begin{figure}
\centering
\resizebox{0.9\columnwidth}{!}{
  \includegraphics{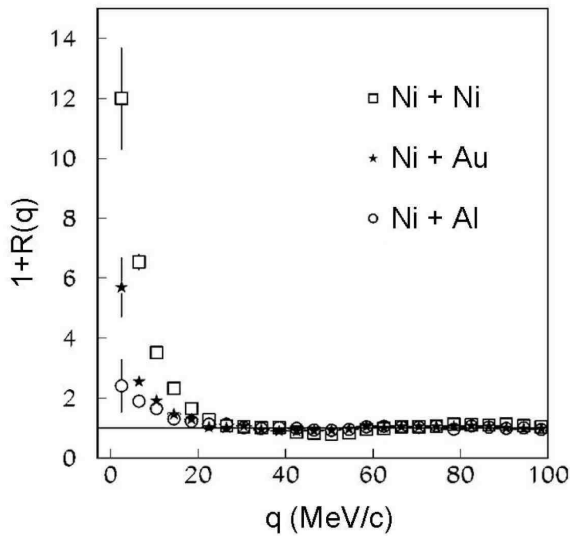}}
  \caption{Two-neutron correlation functions measured in Ni+Ni, Ni+Au and Ni+Al collisions at E/A=45 MeV by the CHIC collaboration \cite{ref46}.}
\end{figure}

The neutron-neutron correlation function is dominated by the Fermionic nature of the neutrons and by their mutual nuclear final state interaction. Unlike the case of proton-proton correlation functions, no Coulomb anti-correlation at small relative momenta exists. The strong maximum at zero relative momentum is caused by the same resonance that brings about the maximum in the $p-p$ correlation function and has been used to investigate the time-scale of the emission and to study the interplay between dynamical and statistical effects \cite{ref40}. Similarly to what was observed in the case of protons, the height of this maximum increases with increasing two-neutron total momentum \cite{ref40,ref45,ref46}, indicating reduced time scales for the emission of more energetic particles emerging from the pre-equilibrium stage. The simultaneous study of $p-p$ and $n-n$ correlation functions in Ni+Al reactions at E/A=45 MeV by means of the Csorgo-Helgesson statistical model (Ref. \cite{ref40} and references therein) has  provided comparable source radii of $R_{G}$=2.7$\pm$0.3 fm for both protons and neutrons. However, slightly larger timescales, with lifetimes of the order of $\tau_{n}\approx$~600$\pm$200~fm/c, appeared to be associated with neutron emission, if compared to protons emitted with lifetimes of about $\tau_{p}\approx$~400$\pm$200 fm/c \cite{ref40}. 

The enhancement at small relative momenta observed in the n-n correlation function in Fig. 11 is due to the short-range nuclear final-state interaction. However, if neutrons are emitted from a long-lived decaying system, the increased average spatial separation between successive nucleons will reduce the significance of the final state strong interaction. The correlation function is then expected to be dominated by quantum statistics effects. An attempt to observe such an effect was successfully performed by the authors of Ref. \cite{ref40a} in the study of compound nuclei formed in $^{18}$O+$^{26}$Mg reactions at E=60 amd 71 MeV. The n-n correlation function was constructed with the evaporated neutrons and a dip at small relative energies was observed. This anti-correlation is generated by the anti-symmetrization of the n-n wave function and represents the first experimental evidence of a pure Fermionic HBT effect. The n-n correlation length is longer than the nuclear interaction length and final interaction effects are negligible.

\begin{figure}
\centering
\resizebox{0.95\columnwidth}{!}{
  \includegraphics{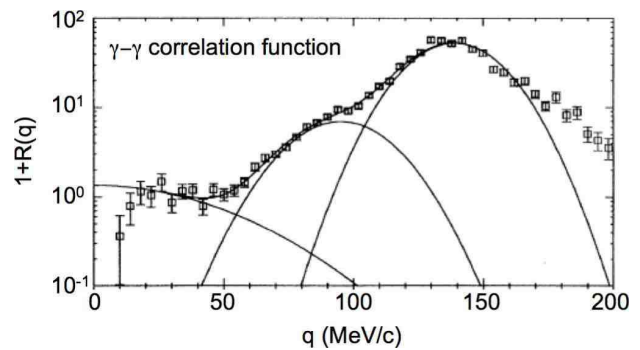}}
  \caption{Data points: Two-photon correlation function measured in Ar+Al reactions at E/A=95 MeV with the MEDEA array \cite{ref50}. The solid lines show calculations of the different effects contributing to the overall shape of the correlation function.}
\end{figure}

Two-photon correlation functions have been measured with the TAPS and MEDEA arrays \cite{ref49,ref50} to extract space-time information about $\gamma$ emitting sources. Figure 12 shows a two-photon correlation function measured in $^{36}$Ar + $^{27}$Al reactions at E/A=95 MeV \cite{ref50}. The pure intensity interferometry effect is observed only at small relative momentum (q$<$45 MeV/c). The peak at 140 MeV/c is due to the neutral pion decay into two photons while the bump at intermediate q-values is produced by badly detected pion decays \cite{ref50}. Because of these complications, two-photon correlation functions were studied with sophisticated analysis techniques yielding Gaussian source sizes of the order of $r_{0}\approx$2 fm and emission lifetimes of the order $\tau\approx$~4~fm/c. These results seem to indicate that $\gamma$-$\gamma$ correlation functions probe mostly the dynamical stage of the reaction. Similar conclusions were also presented in Ref. \cite{ref51} where anticorrelations between energetic $\gamma$-rays and protons were studied with the MEDEA array in Ar+V collisions at E/A=44 MeV. Energetic gamma rays can be considered very good probes of the spatial and temporal properties of the geometric overlap region developed in the very early stages of the collision process when nucleon-nucleon collisions are more important.

\section{Space-time characterization from light complex particles}

During heavy-ion collisions at intermediate energies a large variety of particles and fragments are produced. A complete space-time characterization of the reaction therefore requires multiple intensity interferometry studies extended to several nuclear species. Correlation functions between light charged particles other than protons have indeed been investigated by different authors (Ref.~\cite{ref17} and references therein). 

Figure 13 shows a $d-\alpha$ correlation function measured in the reaction $^{16}$O+$^{197}$Au at E/A=94 MeV \cite{ref52}. The large minimum at small relative momentum is due to the mutual Coulomb repulsion and the peaks at relative momentum $q>30$ MeV/c are due to the nuclear interaction. In particular, the sharp peak around 42 MeV/c corresponds to the first excited state of $^{6}$Li at E*= 2.186 MeV and the broad peak around 84 MeV/c stems mainly from the resonance at E*= 4.31 MeV with small contributions from the resonance at E*= 5.65 MeV. The solid and dashed lines in Fig. 13 show $d-\alpha$ correlation functions calculated from Eq. (4) with the kernel function, $K(q,r)$, constructed from the $d-\alpha$ relative scattering wave function \cite{ref53}.  The best fit of the integral of the first peak at 42 MeV/c provides Gaussian source radii $r_{0}$=4.5 and 3.5 fm for $d-\alpha$ pairs with total energies $E_{\alpha}+E_{d}$~=~55-120 MeV and 120-200 MeV, respectively. As already observed in the case of two-proton correlation functions, the height of the peaks increases with increasing velocity of selected $d-\alpha$ pairs. A closer look at Fig. 13 clearly shows that the whole shape of the $d-\alpha$ correlation function is not reproduced. Eq. (4) does not fit simultanously the first peak at 42 MeV/c and the second peak at 84 MeV/c. The broad peak at 84 MeV/c is better described by using a source radius larger by about 1 fm as compared to the radius extracted from the best-fit of the sharp peak at 42 MeV/c. These difficulties have not been resolved yet and require further research. 

\begin{figure}
\centering
\resizebox{0.9\columnwidth}{!}{
  \includegraphics{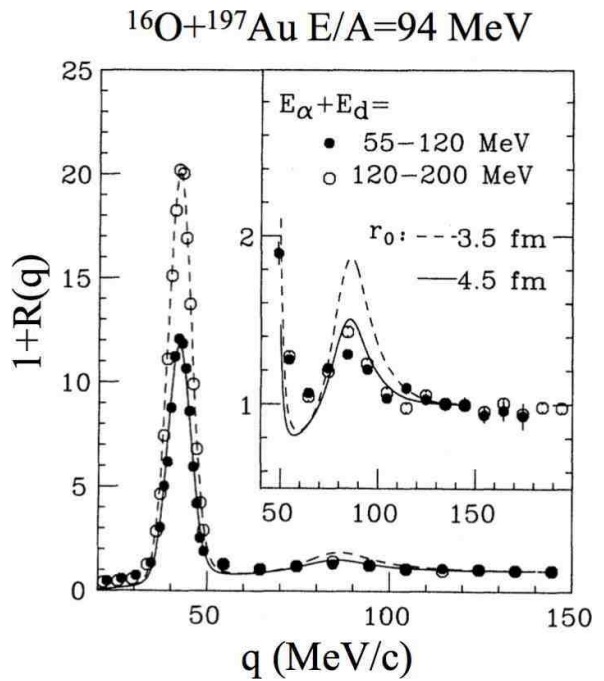}}
  \caption{Data points: Deuteron-alpha correlation functions measured in $^{16}$O+$^{197}$Au reactions at E/A=94 MeV \cite{ref52} for different gates on the total kinetic energy of the particle pairs. The solid and dashed lines correspond to Gaussian analyses of the first peak after taking into account the finite angular resolution of the used experimental setup.}
\end{figure}

Attempts to perform systematic size measurements by using the first sharp peak at q=42 MeV/c can be found in the literature \cite{ref17}. For instance, Fig. 14 shows the comparison between source sizes extracted from $p-p$ correlation functions (left side) and source sizes extracted from the best-fit of the first peak at 42 MeV/c in $d-\alpha$ correlation functions (right side) measured in the same reaction systems, i.e. $^{16}$O+$^{197}$Au at E/A=94 MeV, $^{40}$Ar+$^{197}$Au at E/A=60 MeV and $^{14}$N+$^{197}$Au at E/A=35 MeV (Ref. \cite{ref17} and references therein). Both $p-p$ and $d-\alpha$ source radii decrease with increasing energy of the detected pairs. Furthermore, $d-\alpha$ radii appear to be smaller than $p-p$ source radii. Even if this result seems to be intriguing and requires further research, the difficulties in reproducing the overall shape of the $d-\alpha$ correlation function described in Fig. 13 indicate that these comparisons between $d-\alpha$ and $p-p$ source radii must be taken very cautiously. 

\begin{figure}
\centering
\resizebox{0.92\columnwidth}{!}{
  \includegraphics{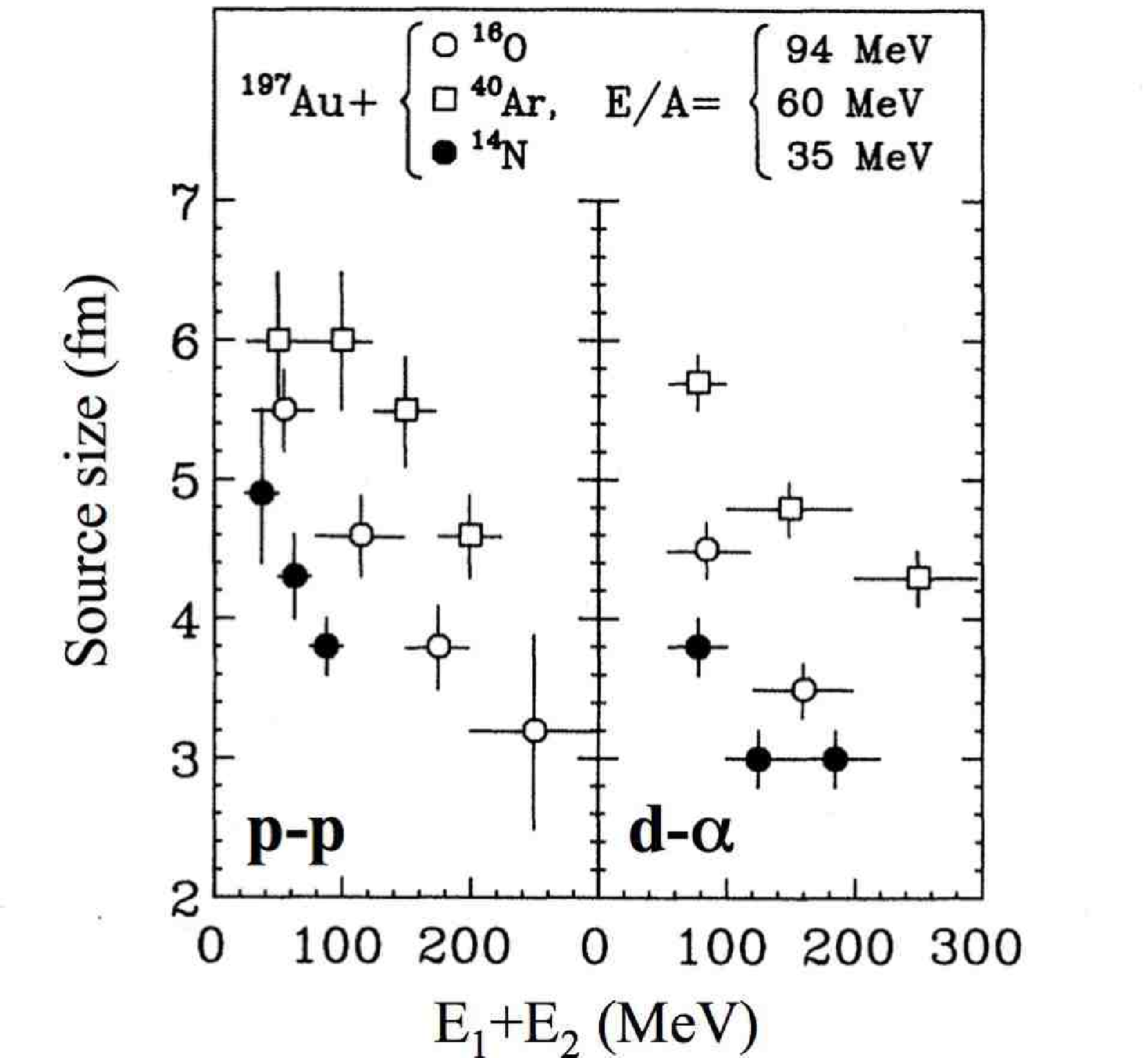}}
  \caption{Comparison between proton-proton (left side) and deuteron-alpha (right side) Gaussian source sizes measured in the same reaction systems and gated on the total kinetic energy of the particle pairs \cite{ref17}.}
\end{figure}

\begin{figure}
\centering
\resizebox{0.92\columnwidth}{!}{
  \includegraphics{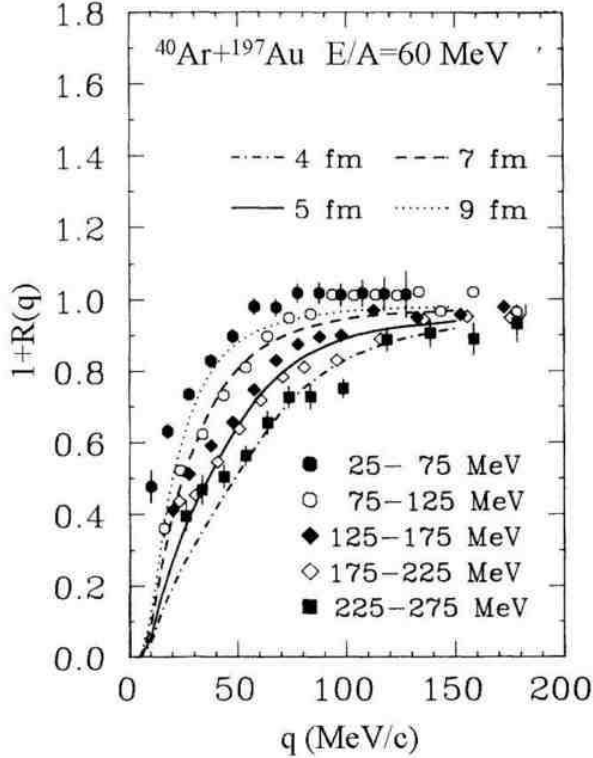}}
  \caption{Deuteron-deuteron correlation functions measured in $^{40}$Ar+$^{197}$Au collisions at E/A=60 MeV \cite{ref54,ref55}. The different symbols correspond to different gates on the total kinetic energy of the detected deuteron pairs. The lines show calculated correlation functions constructed by considering both the Coulomb and the nuclear final-state interactions.}
\end{figure}

Examples of deuteron-deuteron correlation functions measured in Ar+Au at E/A=60 MeV are shown in Fig.~15 \cite{ref54,ref55} for different gates in the total energy of the deuteron pairs (see different symbols). The anti-correlation at small relative momentum, due to the repulsive Coulomb interaction is the main visible feature of these observables. These correlation functions have been analyzed with final state interaction models that include both the Coulomb and the nuclear interaction \cite{ref56,ref57,ref58,ref59,ref60,ref61,ref62}. Source radii from $d-d$ correlation functions have been found to be generally larger than both $p-p$ and $d-\alpha$ radii \cite{ref17,ref57,ref58,ref59,ref60}. 

More recently $d-d$ correlation functions from projectile-like sources produced in Xe+Sn reactions at E/A=50 MeV \cite{ref62} have been measured with the INDRA 4$\pi$ array \cite{ref63,ref64,ref65}. By means of a quantum model including both Coulomb and nuclear final state interactions \cite{ref66}, deuteron emission times from the decay of the projectile-like fragment at different impact parameters were estimated. As the impact parameter decreases from peripheral to more central collisions, the emission times were observed to decrease from 200 fm/c up to 25 fm/c. This result was interpreted as an increasingly important contribution of out-of-equilibrium emission as one moves towards more and more central events where more excited projectile-like fragment are produced. A better angular resolution to measure the $d-d$ correlation function at very low relative momentum \cite{ref62} and an improved knowledge of the $d-d$ final state interaction \cite{ref56} could improve our understanding of deuteron emission mechanisms. 

Triton-triton correlation functions have been also studied by some authors with models including both Coulomb and nuclear final-state interactions or with simplified prescriptions using only the Coulomb repulsion between the coincident tritons \cite{ref57,ref58,ref60,ref61}. The extracted source radii are comparable to those extracted from $d-d$ correlation functions, even if uncertainties in the knowledge of the triton-triton nuclear interaction still exist \cite{ref17}. 

In this subsection we have not discussed the possibility that the correlation function might be distorted by the Coulomb field of the residual system. Only the mutual two-body final state interactions are indeed often used in the literature. This approximation is reasonable only in the case of correlations between particles with identical charge-to-mass ratios causing them to experience a similar acceleration in the Coulomb field of the residual system. Figure 16 shows proton-alpha correlation functions measured in $^{40}$Ar+$^{197}$Au reactions at E/A=60 MeV \cite{ref56,ref67}. The open (solid) points refer to particle pairs with the proton being faster (slower) than the alpha, $v_{p}>v_{\alpha}$ ($v_{p}<v_{\alpha}$). The correlation function displays two peaks at about 15 and 50 MeV/c. The first peak is due to the sequential decay of $^{9}$B, i.e. $^{9}$B $\to$ p+$^{8}$Be $\to$ p+($\alpha$+$\alpha$) with detection of a proton and an $\alpha$ particle only. The second peak at 50 MeV/c is related to the unbound ground state of $^{5}$Li, which has a lifetime of about 130 fm/c. 

\begin{figure}
\centering
\resizebox{0.98\columnwidth}{!}{
  \includegraphics{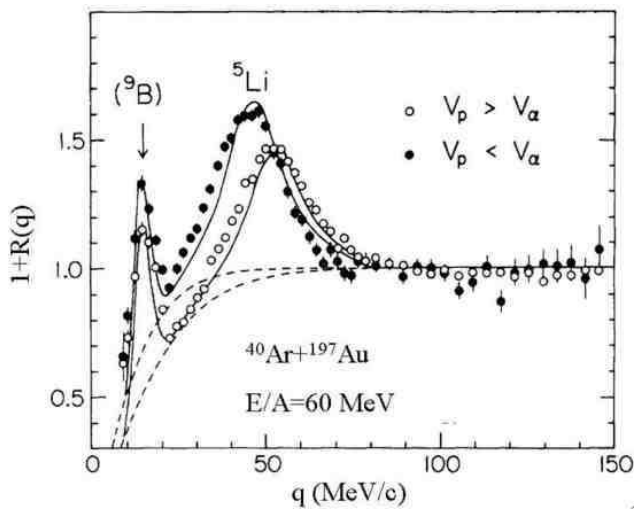}}
  \caption{Proton-alpha correlation functions measured in $^{40}$Ar+$^{197}$Au collisions at E/A=60 MeV \cite{ref56,ref67}. The closed and open circles are obtained by choosing particle pairs in which the proton is, respectively, faster or slower than the alpha particle. }
\end{figure}

While the location of the peak at 15 MeV/c is not altered by the gate on the proton and alpha particle velocities, the second peak is observed at slightly higher relative momenta if pairs with $v_{p}>v_{\alpha}$ are selected. Since the charge-to-mass ratio of protons is greater than that of the alpha particles, the proton will experience a larger acceleration in the Coulomb field of the residual nuclear system. As a consequence, the relative velocity of the pair will be decreased if $v_{p}<v_{\alpha}$ and increased if $v_{p}>v_{\alpha}$, resulting in a shift of the peak position. After taking these distortions into account, $p-\alpha$ correlation functions have been analyzed in $^{14}$N+$^{197}$Au at E/A=35 MeV \cite{ref56,ref58} and the extracted source radii appear to be comparable to the radii obtained from $d-d$ and $t-t$ correlations and larger than those extracted from $d-\alpha$ correlation functions. 

The existing systematics \cite{ref17} shows that two-particle source sizes change when different nuclear species are selected. The obtained differences need further investigations but they all point to the fact that different particles originate from different emitting stages and sources produced during the reaction. The complex scenario emerging from these studies calls for higher resolution detectors and improved theoretical descriptions of final state interaction models for two-body correlations in heavy-ion collisions. In the next sections we will discuss the work that has been done to try to understand time ordering in the emission of light particles. The study of emission chronology is indeed very important in order to improve our understanding of particle production mechanisms in heavy-ion collisions.

\section{Particle emission time and chronology}

Particles originating from one specific source are emitted during a certain time interval. Normally, the rate of emitted particles changes during this time interval, leading to a specific time distribution for the source. If a two-particle correlation function of identical particles could be constructed from particles emitted from a single source (i.e. if the particle source could be isolated), and if the spatial distribution would be known, then the shape of the correlation function would yield information on the shape of the time distribution. Normally, neither the spatial nor the temporal distributions are known. In this case it is not straightforward to extract the spatial and time distributions from the shape of the correlation function. To interpret the experimental results, source models are often used. These source models contain some pre-assumption on the spatial and temporal distributions. The shape of these distributions can, to some extent, be varied by varying parameters of the model (such as radius and temporal width parameters). Such parameters then represent average emission point and average emission time, though it should be remembered that such average values are model dependent. In experimental data the situation is much more complex, since it is never possible to completely isolate one source from all the other sources present during the reaction. The contribution from several sources leads to a total time distribution with a complex shape, where the effective average emission time will depend on different factors. 

What one is usually referring to with {\it emission chronology} is a difference in the average emission times between two particle types. Though, it is worth noticing that the emission times of the two particles may overlap to a large extent, and the difference in their respective average emission times can be small compared with the width of the emission time distributions. Furthermore, if different 
\linebreak[4]
sources contribute, like in peripheral reactions, they contribute to a complex time distribution of the emitted particles. As an example, we can consider two particle types emitted from two sources. Different origins for different average emission times can be recognized: one origin could be a simple shift of two similar time distributions. Another origin could be that the width of the time distributions are different, while their shape is quite similar. Finally, it is also possible to think that, if more than one source is contributing, the relative weight of the sources is different for the two particle types, leading to different average emission times. In a real reaction, all of the above reasons contribute, with a different weight depending for example on applied gatesor selected angular ranges. 

Even though the extraction of the emission times and sequences is quite involved, there is a wealth of experi- mental information that is available. A great advantage is to perform simultaneous measurements of both like and unlike particle correlation functions. By applying different gates, such as polar angle gates, total momentum gates, directional gates, or velocity gates for unlike particles, certain sources are enhanced relative to others. For instance, non-equilibrium emission can be enhanced by high and intermediate total momentum gates, while particles from evaporation and excited fragments can be suppressed by shape analyses disregarding the very low relative momentum region. Furthermore, single particle information, such as energy spectra at different angles, should be used in the analysis. A systematic study can, therefore, to a large extent disentangle the space-time characteristics of the contributing sources, thereby putting strong constraints on theoretical models. 

\subsection{Emission chronology from like particle correlations}

The emission chronology between two particle types can, under certain conditions, be determined from like-particle correlation functions. If it is valid to assume that both particle types are emitted from the same spatial region, a fit with a calculated correlation function, based on some source function, to the experimental correlation function, will yield an average emission time for each particle type. By comparing these average emission times, an emission chronology can be inferred. For a review, see Ref. \cite{ref68} and references therein. The same method can also be used to determine differences in emission times for the same particle type emitted from different systems but with the same spatial region, for instance, different neutron emission times from systems similar in size and energy content, but with different isospin content (see next section). 

The drawback of this method is that the results are sensitive to the assumption of emission from the same spatial region. Furthermore, the extracted average emission times are model dependent, since the average emission times depend on the pre-assumption of the shape of the (spatial and) temporal distributions assumed by the specific source model. 

\subsection{Emission chronology from unlike particle correlations}

Model independent information on the emission chronology of two particle types can instead be obtained from unlike-particle correlation functions. If there is a difference in the average emission times, it is possible, by suitable gates, to divide the particle pairs into two classes with different average distances when the two particles "start to interact". Since the strength of the final state interaction depends on the distance between the particles, this will lead to a different strength of the correlation function for the two classes. By comparing these correlation functions, the emission chronology can then be inferred without any model assumptions. 

Below we present the details of this technique. An important assumption for this method to be valid is that the particles are emitted independently. Furthermore, certain assumptions (which depends on the used gates) must be  made on the spatial region from which the particles are emitted.

\subsubsection{Velocity gated correlation functions}

A technique to probe the emission sequence and time delay of ejectiles in nuclear reactions was first suggested for charged particle pairs, based on the idea that mutual Coulomb repulsion would be experienced by pairs of charged particles emitted with a short time delay. Comparison of the velocity difference spectra with trajectory calculations would thus give a measure of the average particle emission sequence \cite{ref69,ref70,ref71}. 

The technique was extended to any kind of interacting, non-identical particle pairs in the theoretical study of Ref. \cite{ref72}. There it was demonstrated that the sensitivity of the correlation function to the asymmetry of the distribution of the relative space-time coordinates of the particle emission points can be used to determine the differences in the mean emission times by applying energy or velocity gates. This effect has been proposed for particle pairs such as $pd$ and $np$ \cite{ref72}, $p\pi$ \cite{ref73}, $K^{+}K^{-}$ \cite{ref74}. 

Velocity gated correlation functions of non-identical particles is a very powerful tool to investigate emission sequences in nuclear collisions \cite{ref75}. The basic idea is that, if there is an average time difference in the emission times of two particles types, there will also be a difference in the average distance for particle pairs selected with the condition $v_{1}>v_{2}$ as compared to the pairs selected with the complementary condition $v_{1}<v_{2}$. This is because the particle emitted first will, on average, travel a different distance in the two complementary classes (due to the different velocities) before the second particle is emitted. In particular, the interaction is enhanced for those pairs for which the average distance is smaller. This can be easily seen if one compares the correlation function $C_{1}(q)$, gated on pairs $v_{1}>v_{2}$, with the correlation function $C_{2}(q)$, gated on pairs $v_{1}<v_{2}$. If particle 2 is emitted earlier (later) than particle 1, than the condition $v_{1}>v_{2}$ will sample smaller distances (larger correlations) than the complementary condition $v_{1}<v_{2}$. Therefore the ratio $C_{1}/C_{2}$ will show a peak (dip) in the region of relative momentum q where there is a correlation and a dip (peak) where there is an anti-correlation. Furthermore the ratio $C_{1}/C_{2}$ will approach unity both for $q\to 0$ (since the velocity difference of the two emitted particles is negligible) and $q\to\infty$ (since modifications of the two-particle phase space density arising from final state interactions are negligible). The exact location of the peak and dip in the ratio depends on the source, and in particular on the origin of the difference in the average emission times.

\subsubsection{Experimental results}

\begin{figure*}
\centering
\resizebox{1.4\columnwidth}{!}{
  \includegraphics{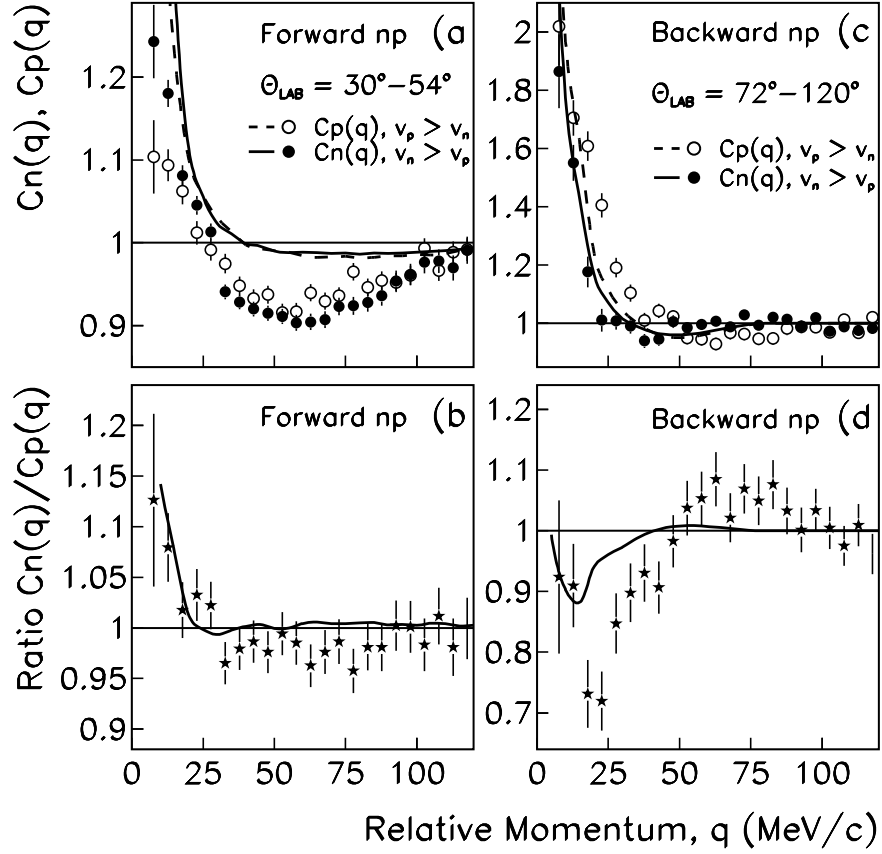}}
  \caption{From the E/A=61 MeV $^{36}$Ar+Al reaction measured at KVI, velocity-gated $np$ correlation functions ($C_{n}$, filled circles, and $C_{p}$, open circles) and their ratio ($C_{n}/C_{p}$) Left column: forward measurement. Right column: backward measurement.  The lines are results from model calculations. From Ref. \cite{ref77}.}
\end{figure*}

At low energies, the analysis of two-particle correlations and velocity difference spectra has allowed to find results consistent with the statistical compound nucleus decay for the light products emitted from the reactions $^{16}$O+$^{10}$B (E$_{lab}$=62.5 MeV), and $^{16}$O+$^{12}$C (64 MeV). The method could also be used for the determination of fission time scales \cite{ref76}. 

At intermediate energies, the time sequence of p and d has been deduced for the E/A=50 MeV Xe+Sn reaction studied by the INDRA collaboration. An average emission of deuterons $\approx$ 250 fm/c earlier than protons has been explained as the result of averaging over a long time sequence between pre-equilibrium and thermal emission for protons, whereas deuteron emission, resulting mainly from hard nucleon-nucleon collisions, is concentrated at a few tens of fm/c \cite{ref62}. 

First experimental evidence of the emission chronology of neutrons and protons deduced from the $np$ correlation function, was reported in Ref. \cite{ref75} for the E/A=45 MeV $^{58}$Ni+$^{27}$Al reaction measured by the CHIC collaboration at LNS, Catania. The experimental results from differently gated correlation functions were in qualitative agreement with the Koonin-Pratt formalism \cite{ref23,ref24}. It was claimed that for events, selected for high-parallel-velocity and high-total-momentum which enhances projectile residue and/or intermediate velocity sources, the proton is, on average, emitted earlier than the neutron. 

In Ref. \cite{ref77} the emission time chronology of neutrons, protons, and deuterons, was presented for the E/A=61 MeV $^{36}$Ar+$^{27}$Al reaction measured at KVI, Groningen. The experimental results showed that the angular and total-momentum dependences of the $pp$ and $np$ correlation functions support a dissipative binary reaction scenario, where early dynamical emission is followed by statistical evaporation. The reverse kinematics utilized in the experiment, and fairly high energy thresholds, enhanced the early dynamical emission component in the backward measurement. The analysis of velocity-gated correlations of non-identical particle pairs yielded detailed information about the particle emission time sequence. The results from the $np$ backward measurement are shown in Fig. 1, right column (from Ref. \cite{ref77}). The dip in the $C_{n}/C_{p}$ ratio indicates that neutrons are, on the average, emitted earlier than protons. For the forward measurement (Fig. 17, left column), the shape of the correlation function exhibits a correlation at $q<$40 MeV/c and a small anti-correlation at 40$<q<$100 MeV/c, and the pairs with $v_{n}>v_{p}$ contributing to $C_{n}$ interact more strongly. This indicates that protons are, on the average, emitted earlier than neutrons, a result in agreement with that obtained for the $\theta_{lab}\approx$~45$^{o}$ E/A=45 MeV $^{58}$Ni+$^{27}$Al reaction \cite{ref75}. The complete sequence of average emission times, $\tau$, extracted from the E/A=61 MeV $^{36}$Ar+$^{27}$Al reaction was the following: for the dynamical emission source, $\tau_{n}<\tau_{d}<\tau_{p}$; for the projectile residue emission, $\tau_{d}<\tau_{p}<\tau_{n}$. The interpretation of these results, presented in Ref. \cite{ref77}, highlights the importance of the contribution from the different emission sources.

The technique of Ref. \cite{ref72} has also been applied to pairs of non-identical light charged particles produced in central collisions of heavy ions in the A=100 mass region at a beam energy of 400 MeV/nucleon, measured with the FOPI detector system at GSI \cite{ref78}. The difference between longitudinal correlation functions with the relative velocity parallel and anti-parallel to the center-of-mass velocity of the pair in the central source has allowed the extraction of apparent space-time differences of the emission of the charged particles. Comparing the correlations with results of a final-state interaction model, delivered quantitative estimates of these asymmetries. Time delays as short as 1 fm/c or - alternatively - source radius differences of a few tenth fm were resolved. The strong collective expansion of the participant zone introduces not only an apparent reduction of the source radius but also a modification of the emission times. After correcting for both effects a complete sequence of the space-time emission of $p$, $d$, $t$, $^{3}$He, $\alpha$ particles was extracted. 

At even higher energy regimes, the above method has been used to tackle the problem of the possible observation of strangelets in the frame of the distillation process following the creation of a quark gluon plasma. In this case, strange and anti-strange particles may not be produced at the same time in a baryon rich system under low bag constant scenarios. Such a prediction has been tested using $K^{+}K^{-}$ correlations in Ref. \cite{ref79}.

The knowledge of the emission times and chronology is very important in
order to extract information on the nuclear interaction from complex
nuclear collisions. It is, however, not straightforward to obtain emission 
times and chronology from a nuclear reaction, and different methods
need to be combined. Using velocity gates is a relativeley new and promising
method to, model independently, extract the emission chronology. Examples
of first experimental results using velocity gates have been presented,
together with their interpretations. To exclude other explanations, the
method needs to be combined with other probes and methods to draw the
correct conclusions. This is a field where further developments can
be made.

\section{Isospin effects and perspectives for the asy-EOS} 

\begin{figure*}
\centering
\resizebox{1.55\columnwidth}{!}{
  \includegraphics{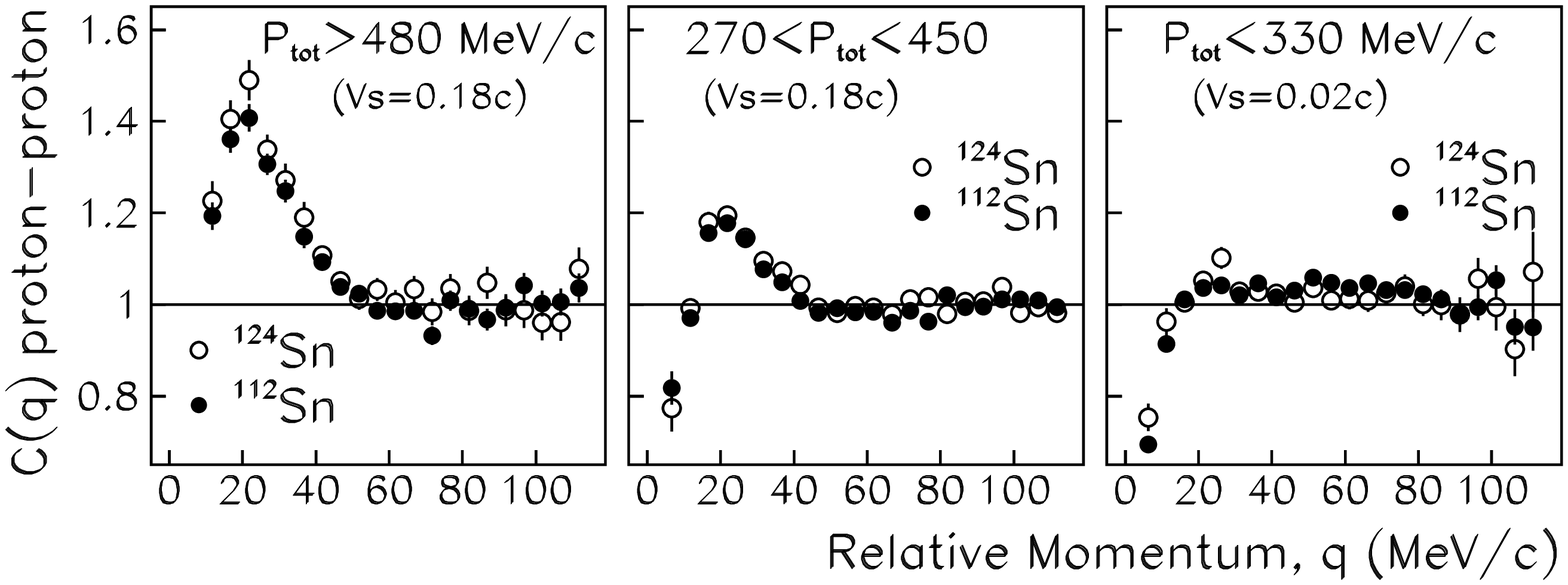}}
  \caption{For $^{36}$Ar+$^{112}$Sn (filled circles) and $^{36}$Ar+$^{124}$Sn (open circles) collisions, $pp$ correlation functions. Left: $pp$, high-$P_{tot}$, 30$^{o}\leq\theta\leq$42$^{o}$. Middle: $pp$, intermediate-P$_{tot}$, 30$^{o}\leq\theta\leq$42$^{o}$. Right: $pp$, low-P$_{tot}$, 54$^{o}\leq\theta\leq$114$^{o}$. From Ref. \cite{ref85}. }
\end{figure*}

\begin{figure*}
\centering
\resizebox{1.55\columnwidth}{!}{
  \includegraphics{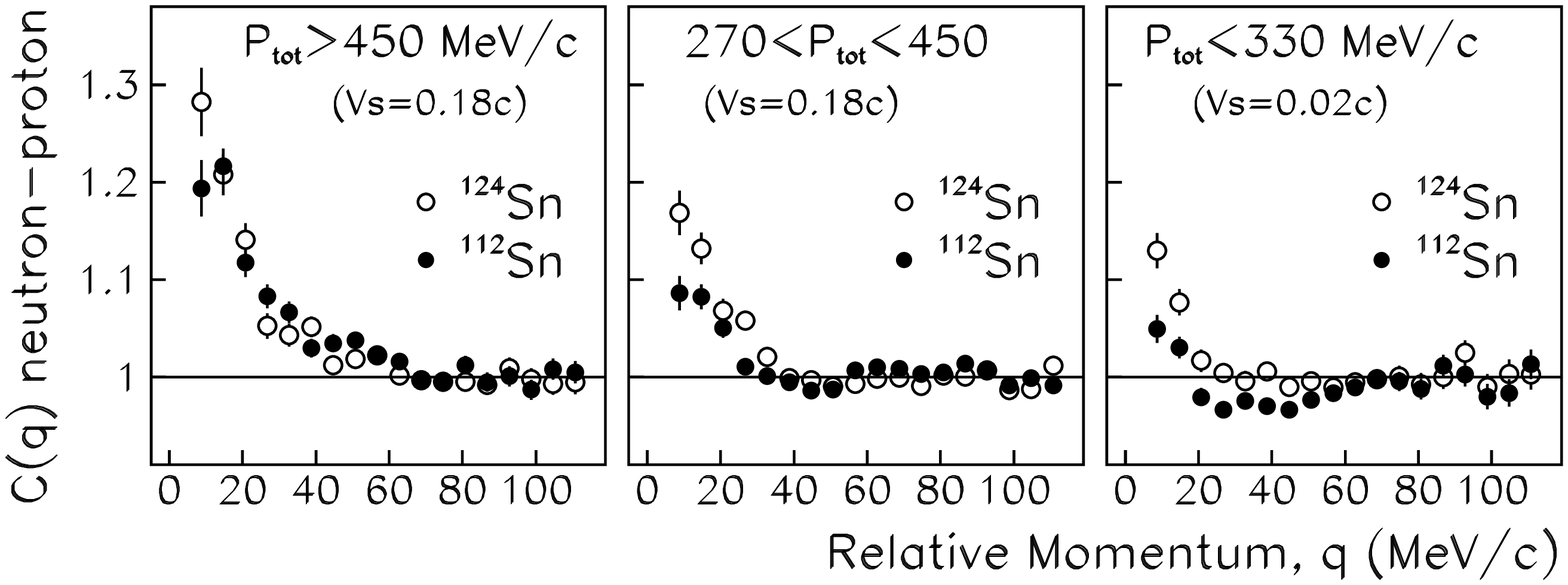}}
  \caption{For $^{36}$Ar+$^{112}$Sn (filled circles) and $^{36}$Ar+$^{124}$Sn (open circles) collisions, $np$ correlation functions. Left: $np$, high-P$_{tot}$ , 30$^{o}\leq\theta\leq$120$^{o}$. Middle: $np$, intermediate-P$_{tot}$, 30$^{o}\leq\theta\leq$120$^{o}$. Right: $np$, low-P$_{tot}$, 54$^{o}\leq\theta\leq$120$^{o}$. From Ref. \cite{ref85}. }
\end{figure*}

The isospin dependence of the nuclear equation of state is probably the most uncertain property of neutron-rich matter. This property is essential for the understanding of extremely asymmetric nuclei and nuclear matter as it may occur in the $r$-process of nucleosynthesis or in neutron stars. In order to study the isospin-dependent EOS, heavy ion collisions with isotope separated beam and/or target nuclei can be utilized. In these collisions, excited systems are created with varying degree of proton-neutron asymmetry and a noticeable isospin dependence of the decay mechanism is expected \cite{ref80}. By selecting semi-peripheral collisions, the symmetry term of the EOS is expected to be probed at low densities, while central collisions are expected to be sensitive to the high-density dependence. 

\subsection{Theoretical predictions}

Recently, the two-nucleon correlation function has been considered as a probe for the density dependence of the nuclear symmetry energy \cite{ref81,ref82,ref83}. In these theoretical studies with an isospin-dependent IBUU transport model, it was shown that the density dependence of the symmetry term of the EOS affects the temporal and spatial structure of reaction dynamics by affecting the average emission times of neutrons and protons as well as their relative emission sequence. For central collisions, a stiff EOS causes high momentum neutrons and protons to be emitted almost simultaneously, thereby leading to strong correlations. A soft EOS delays proton emission, which weakens the $np$ correlation \cite{ref81,ref82}. It was found that the symmetry energy effect becomes weaker with increasing impact parameter and incident energy. Also, the strength of the nucleon-nucleon correlation function is reduced in collisions of heavier reaction systems as a result of larger nucleon emission source \cite{ref82}. It was further found that the momentum dependence of both isoscalar nuclear potential and the symmetry potential influences significantly the space-time properties of the nucleon emission source. Specifically, the momentum dependence of the nuclear potential reduces the sensitivity of two-nucleon correlation functions to the stiffness of the nuclear symmetry energy \cite{ref83}. 

In spite of the large uncertainties in the symmetry potential, and in particular in the momentum dependent part, the exploratory studies in \cite{ref81,ref82,ref83} are very encouraging for using two-particle correlations to study the symmetry potential. It is also important to remember that the symmetry interaction does not only influence the dynamical emission of particles from the overlap and neck-like regions, but also the formation of the residues. Therefore also the particles emitted from the residues contain information on the symmetry energy. To make improvements in the future on the understanding of the symmetry potential, it is necessary to have models that consistently can describe both the dynamical emission of particles and the formation of residues and their subsequent pre-equilibrium and equilibrium emission of particles. By applying the experimental conditions and systematically comparing calculated and experimental energy spectra and gated correlation functions for like and unlike particles, it will be possible to obtain hard constraints on the symmetry potential.

\subsection{Experimental results}

First experimental results have been obtained on two-particle correlation functions from systems similar in size, but with different isospin content \cite{ref84}. Small-angle two-particle correlation functions with neutrons and protons have been obtained from semi-peripheral E/A=61 MeV Ar+$^{112,124}$Sn collisions measured at the AGOR cyclotron of KVI. 

The emission from the different sources was enhanced or suppressed by introducing angular cuts (intermediate-velocity-source emission is enhanced at forward, and target-residue-emission at backward angles) and cuts in the total momentum ($P_{tot}$) of the particle pair, calculated in the relevant emission source frame. Figure 18 (from Ref. \cite{ref85}) presents the pp correlation function for particle pairs selected within the three different gates.

\begin{enumerate}
\item
Particles emitted by the intermediate velocity source at prompt dynamical emission stage, (e.g. first-chance nucleon-nucleon collisions), are enhanced by selecting high-$P_{tot}$ pairs in the intermediate velocity source 
\linebreak[4]
frame. For the sample of $pp$ pairs, the angular range 30$^{o}\leq\theta\leq$~42$^{o}$ is used for this gate.
\item 
Particles emitted by the intermediate velocity source at a later dynamical emission stage (e.g. neck-emission), are enhanced by selecting intermediate-$P_{tot}$ pairs in the intermediate velocity source 
frame. Again, the angular range 30$^{o}\leq\theta\leq$~42$^{o}$ is utilized for $pp$ pairs.
\item
Particles emitted by the target residue are enhanced by selecting low-$P_{tot}$ pairs in the target residue frame. The angular range 54$^{o}\leq\theta\leq$~120$^{o}$ is applied to both $pp$ and $np$ pairs. The neutron energy threshold is set to 8 MeV for this gate. 
\end{enumerate}

One can notice that the height of the peak at q$\approx$20 MeV/c is progressively reduced going from gate 1 to gate 3, indicating an increase in the particle emission time. Figure 19 presents the corresponding results for the $np$ correlation function. By comparing the results for the two Sn targets, one can see an isospin effect, particularly in gate 1 for $pp$ and in gates 2 and 3 for $np$, where the height of the correlation function is larger for the more neutron-rich target. This indicates a shorter average emission time for this system. 

For the interpretation of the correlation data, it is important to note that the correlation function depends on the space-time extent of the emitting source. From the size of the source, a stronger correlation is expected for the smaller $^{36}$Ar+$^{112}$Sn system, an effect expected also because of the larger excitation energy per particle available for this system (yielding a shorter emission time). On the other hand, the change in neutron number implies a different symmetry energy which also affects the n (and p) emission times. Neutrons are expected to be emitted faster in the neutron-rich system, which would lead to an enhancement of the correlation strength for $^{36}$Ar+$^{124}$Sn. Thus, the net influence on the correlation function is not easily predictable, both due to the uncertainty in the symmetry energy and to the presence of more than one source of emission. The stronger $np$ correlation observed for the larger Ar+$^{124}$Sn system in the low-total-momentum gate may indicate that the more asymmetric system generates a more asymmetric and excited target residue that, consequently, decays on a faster time scale. 

More insight into these results has been gained by performing an analysis of the particle emission time sequence in Ref. \cite{ref86}. In all studied angle and total-momentum gates, it was found that neutrons are, on average, emitted earlier than protons. Furthermore, the shorter $np$ emission time scale for the Ar+$^{124}$Sn system results from a faster emission of the neutrons, as compared to the Ar+$^{112}$Sn system. This is particularly true for the particles emitted from the target residue, indicating that the residues in the two reactions were formed differently due to the symmetry interaction. Further experimental results of particle emission sequence involving deuterons indicate that for the Ar+$^{124}$Sn system neutrons are emitted slightly earlier than deuterons, again a result pointing to a faster neutron emission for the more neutron rich system. Deuterons, being formed mainly by coalescence, appear to have emission times that fall in between that of neutrons and protons. No sizeable isospin effects in the emission sequence of deuterons and protons were found for the above mentioned systems.

\section{Accessing the space-time properties at freeze-out} 

IMF-IMF correlation functions are expected to provide the space-time properties of nuclear systems produced in nuclear reactions at the time when they approach the freeze-out stage. This can offer a unique opportunity to better understand the mechanisms of multifragmentation phenomena and their possible links to a liquid-gas phase transition in nuclear matter. 

IMF-IMF correlation functions are usually constructed by combining together fragments with different charges and masses into a single correlation function. This is accomplished by sorting all IMF-IMF pairs with respect to the so-called reduced velocity, $v_{red}=v_{rel}/\sqrt{Z_{1}+Z_{2}}$, where $Z_{1}$ and $Z_{2}$ are the charges of the two IMFs \cite{ref87,ref88,ref89,ref90,ref91,ref92,ref93,ref94}. Typical measured correlation functions are shown in Figs. 20-24 and 26. 

Theoretically, final state interaction models including only the Coulomb repulsion have often been used to analyze IMF-IMF correlation functions. Indeed, fragment spatial separations are on average expected to be larger than the range of the nuclear force. This approximation needs certainly to be verified within the context of more quantitative models. However, the most prominent feature observed in IMF-IMF correlation functions is represented by the strong Coulomb anti-correlation at small relative velocities (see Figs. 20-24,26). Possible resonances corresponding to the decay of heavier fragments either occur at higher relative velocities not easily accessed experimentally or require higher energy and angular resolution in order to be clearly resolved. Such limitations limit the study of IMF-IMF correlation functions by means of simplified Coulomb interaction models. The IMF-IMF Kernel function in Eq. (4) has been calculated either from the relative Coulomb wave function  or by using a classical Coulomb treatment leading to the simple expression, $K(r,q)=\left(1-r/r_{c}\right)^{1/2}-1$, with $r_{c}=2\mu Z_{1}Z_{2}e^{2}/q^{2}$  and $Z_{1}$ and $Z_{2}$ being the charges of the two coincident fragments \cite{ref87}. The emitting source function is commonly assumed to have a spherical shape with radius $R_{S}$ and it is assumed to be characterized by a temperature $T$. Fragments are then emitted according to an exponential time profile, $P(t)\propto\exp\left(-t/\tau\right)$. The radius, $R_{S}$, and the lifetime, $\tau$, are deduced as free parameters from the best fit of the experimental correlation functions, thus providing the space-time characterization of IMF emission. Figure 20 shows correlation functions calculated with sources having emission lifetimes between 50 and 500 fm/c. It can be observed that the shape of the correlation function at small reduced velocities ($V_{red}<$30$\cdot$10$^{-3}$c) is strongly affected by fragment emission times. The calculations in Fig. 20 are performed by using both the classical (lines) and the quantum (dots) two-body kernel function in Eq. (4), providing virtually identical results \cite{ref87}. The width of the Coulomb anticorrelation at small reduced velocities is used to extract IMF emission times.

\begin{figure}
\centering
\resizebox{0.95\columnwidth}{!}{
  \includegraphics{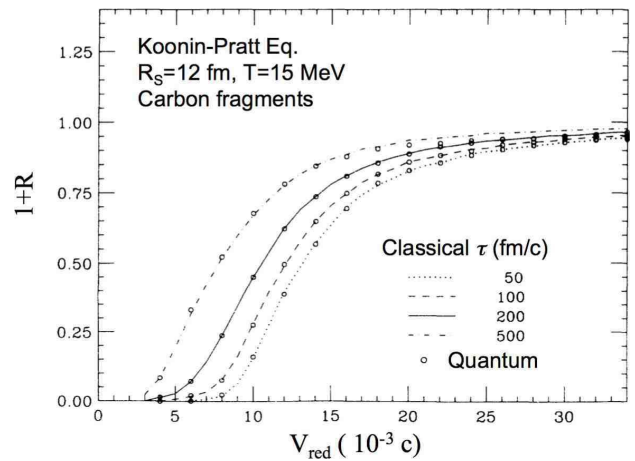}}
  \caption{IMF-IMF correlation functions calculated for different emitting source lifetimes \cite{ref87}. The lines and dots correspond, respectively, to classical and quantum calculations.}
\end{figure}

\begin{figure}
\centering
\resizebox{0.95\columnwidth}{!}{
  \includegraphics{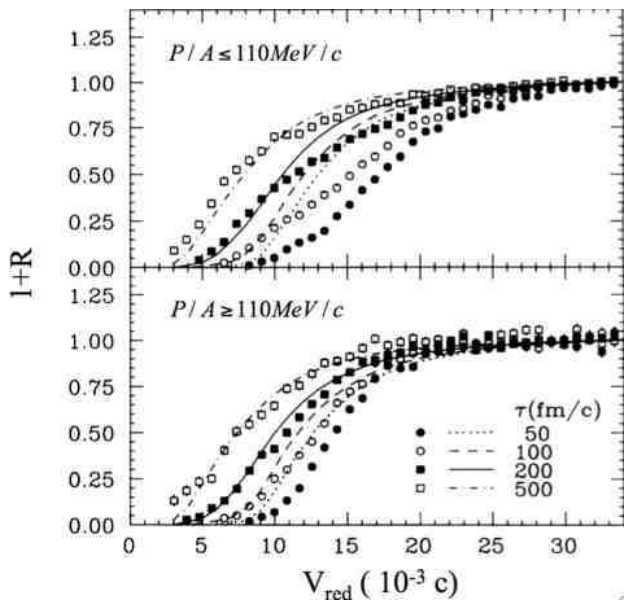}}
  \caption{Two-body (lines) and three-body (dots) trajectory calculations for IMF pairs with total momenta per nucleon, P/A$\leq$110 MeV/c (top panel) and P/A$\geq$110 MeV/c (bottom panel). The calculations are performed for different values of the IMF emission lifetimes. }
\end{figure}

Alternative 3-body approaches to IMF-IMF correlation functions are based on trajectory calculations where two IMFs are emitted sequentially with some average time delay, and their motion is propagated to the detectors under the influence of both their mutual final state Coulomb interaction and the repulsion induced by the Coulomb field of the residual system \cite{ref88,ref89,ref90}. The correlation function is then deduced from Eq. (2) applied to the simulated coincidence and single-particle yields \cite{ref88,ref89,ref91}. Figure 21 is taken from Ref. \cite{ref87} and shows the comparison between two-body classical calculations (lines) and three-body Coulomb trajectories calculations (points) for a source having charge $Z_{S}=$~93, radius $R_{S}=$~12~fm, temperature $T=$~15~MeV and emission lifetimes varying between $\tau=$~50~fm/c and $\tau=$~500~fm/c. The upper and lower panels show the results for carbon pairs emitted with total momenta, $P/A\leq$~110 MeV/c and $P/A\geq$~110 MeV/c, respectively. The use of a very large mass number for the source ($A_{S}=$~10000) ensures that recoil effects are not taken into account and only the effects of the residual Coulomb field are explored. For low momentum gates, fragments are emitted with small initial velocities and distortions in the Coulomb field of the source are large. Indeed, significant differences between three- and two-body calculations exist up to $\tau$=200 fm/c. For longer emission lifetimes, the two-body approach still remains valid. As it can be easily expected, in the case of fragment pairs at high total momentum (lower panel), fragments move so quickly that the effects induced by the Coulomb field of the residual system acting as a third body can be neglected. The two-body approach is found to be still reasonable if the emission times are longer ($\tau>200$ fm/c). For shorter emission times, the disagreement is substantial and three-body correlations seem to be important, even if the effect is less pronounced than in the case of low momentum fragments (upper panel in Fig. 21). Three-body trajectories and even more realistic N-body Coulomb trajectory calculations have extensively been used in the literature \cite{ref88,ref92,ref93,ref94}. 

More recently IMF-IMF correlation functions have also been compared to predictions of microscopic models where both the Coulomb and the nuclear interactions are taken into account \cite{ref95,ref96}. Microscopic models provide a more realistic approach and represent certainly a promising opportunity for the future in order to have a clear and unambiguous space-time characterization of complex fragment emission mechanisms.

 \section{Characterizing multifragmentation phenomena} 
 
 Understanding whether multifragmentation results from a sequence of binary splittings or rather from a simultaneous break-up of an excited nuclear system has certainly represented one of the main questions raised in the last two decades. 
A sequential binary splitting would correspond to cluster emission from the surface of an excited source (similar to fission). 
 This process is associated with long emission times of the order of 10$^{-20}$-10$^{-21}$ sec, necessary for shape deformation. In contrast, if multifragmentation corresponds to a simultaneous breakup of nuclear matter, 
the system is expected to fall apart over shorter times (10$^{-22}$-10$^{-23}$ sec), comparable to the timescales involved in the growth of density fluctuations in the spinodal instability region of the nuclear phase diagram. In the following part of this section we will present some of the main results that can be found in the literature on the study of fragment emission times in heavy-ion collisions.
 
\begin{figure}
\centering
\resizebox{0.85\columnwidth}{!}{
  \includegraphics{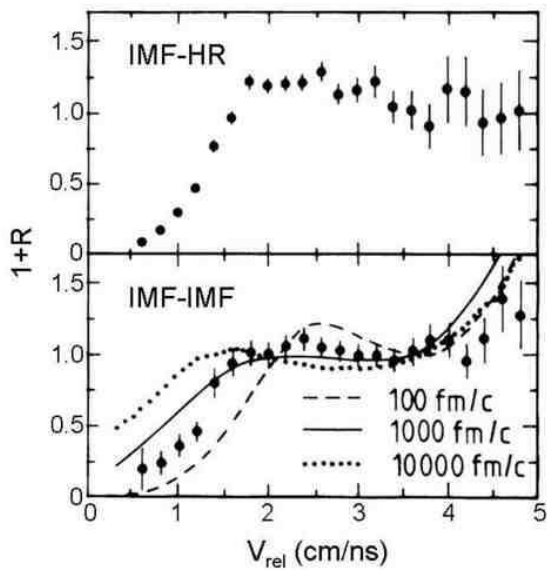}}
  \caption{ Bottom panel: The data points show the IMF-IMF correlation function measured in $^{18}$O+$^{nat}$Ag reactions at E/A=84 MeV \cite{ref88} in events where at least two IMS fragments are detected. The lines correspond to calculations performed with three-body Coulomb trajectories assuming different IMF emission lifetimes. The top panel refers to events where only one IMF and one heavy residues (HR) is observed and shows their correlation function.}
\end{figure}
 
One of the earliest studies of fragment-fragment correlation functions was presented in Ref. \cite{ref88} where the authors have studied $^{18}$O+$^{197}$Au and $^{18}$O+$^{nat}$Ag reactions at E/A=84 MeV. The IMF-IMF correlation function in events where two or more IMFs were observed (Fig. 22, bottom panel), was compared to the correlation function IMF-HR (IMF-Heavy Residue, Fig. 22 top panel) in events where only one IMF and one heavy fragment were observed. The lines in the bottom panel of Fig. 22 show the results of three-body Coulomb trajectory calculations. IMF emission times of the order of $\tau$=1000 fm/c were obtained in both classes of events selected by the authors. These results suggest that the production of more than two IMFs in the studied reactions is characterized by a mechanism similar to the evaporation of one IMF by a heavy fragment. This result supports a sequential binary evaporation mechanism as responsible for IMF emission.

A similar analysis was performed also in Ref. \cite{ref90} where incomplete fusion reactions with the production of two or three heavy (Z$\geq$10) fragments in $^{22}$Ne+$^{197}$Au collisions at 60 MeV/u were investigated. The analysis of relative velocities and angles in the center-of-mass of the coincident fragments was found to favour a sequential emission of fragments with short time steps between two consecutive binary decays \cite{ref90}. 
 
\begin{figure}
\centering
\resizebox{0.9\columnwidth}{!}{
  \includegraphics{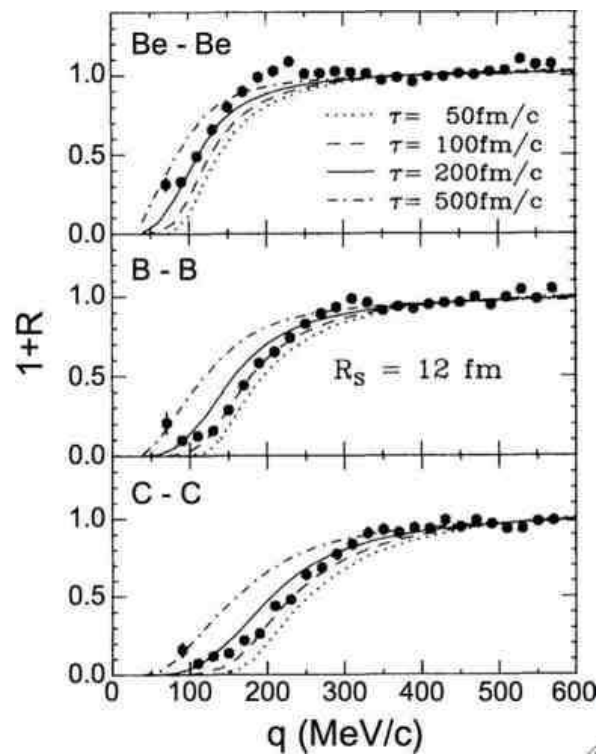}}
  \caption{Data points: beryllium-beryllium (top panel), boron-boron (middle panel) and carbon-carbon (bottom panel) correlation functions measured in Au+Au central collisions at E/A=35 MeV \cite{ref89}. Lines: calculations performed with Eq. (4) using a classical approzimation to the kernel function and assuming different emission lifetimes and a spherical source size of 12 fm.}
\end{figure}

\begin{figure*}
\centering
\resizebox{1.5\columnwidth}{!}{
  \includegraphics{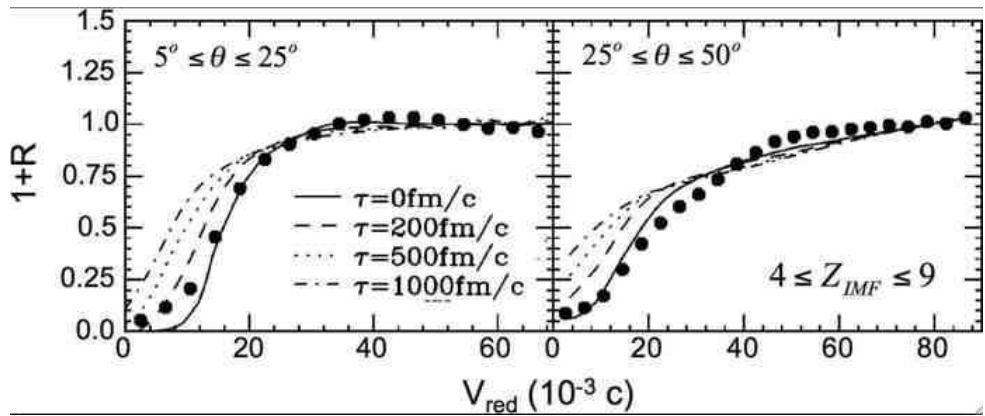}}
  \caption{Data points: IMF-IMF correlation functions measured in central collisions $^{86}$Kr+$^{93}$Nb at E/A=50 MeV at polar angles 5$^{o}\leq\theta\leq$~25$^{o}$ (left panel) and 25$^{o}\leq\theta\leq$~50$^{o}$ \cite{ref93}. N-Coulomb trajectory calculations are shown by lines and correspond to different emission lifetimes for the detected fragments.}
\end{figure*}

The data points in Fig. 23 show Be-Be, B-B and C-C correlation functions measured in central Ar+Au collisions at E/A=35 MeV at polar angles 12$^{o}<\theta<$~35$^{o}$ \cite{ref89}. This angular region allowed the authors to select IMFs produced from incomplete fusion mechanisms (Ref. \cite{ref89} and references therein). The curves in Fig. 23 represent calculations performed with Eq. (4) using a classical approximation for the two-fragment Coulomb kernel function \cite{ref89}. IMFs were assumed to be emitted from a spherical source of radius $R_{S}$=12 fm and with emission times distributed according to an exponential law, $P(t)\propto\exp\left(-t/\tau\right)$. The comparison to experimental data suggests that fragments are emitted with time-scales of about $\tau\approx$~100-200 fm/c. Comparable emission times were obtained with three-body Coulomb trajectory calculations where the IMFs move towards the detectors under the influence of both their mutual final-state Coulomb interaction and the repulsion induced by the Coulomb field of the residual system \cite{ref89}. The obtained emission times are shorter than the timescales characteristic of evaporation processes from compound nuclei. An analysis of total-momentum-gated IMF-IMF correlation functions in the same set of data showed that high energy IMFs are produced with even shorter emission times, of the order of 50 fm/c, thus indicating that fragment emission in central collisions begins in early stages of the reaction and continues throughout the later equilibrium stages \cite{ref89,ref90}. 

The reaction system $^{36}$Ar+$^{197}$Au has also been studied at incident energies, E/A=50 MeV, using an N-body Coulomb approach \cite{ref92}. Angle-averaged correlation functions contain space-time ambiguities that are difficult to resolve: the wide minimum of the correlation function at small reduced velocities cannot be associated with a unique combination of source size, $R_{S}$, and emission time, $\tau$. A long-lifetime emission produces emitting sources elongated in the longitudinal direction defined by the total momentum vector, $\vec{P}$.  Similarly to the case of $p-p$ correlation functions already discussed previously, IMF-IMF directional correlation functions were used to extract emission times, $\tau=$~50 fm/c, from the surface of a dilute source having a density of about $\rho/\rho_{0}\approx$~0.4 \cite{ref92}. 

Central collisions between nearly symmetric systems, $^{86}$Kr+$^{93}$Nb, at E/A=50 MeV were studied in Ref. \cite{ref93} with the goal of disentangling instantaneous from sequential multifragmentation break-up scenarios. Figure 24 
\linebreak[4]
shows correlation functions constructed with IMF fragments having charges 4$\leq Z_{IMF}\leq$9  detected at polar angles 5$^{o}\leq\theta_{lab}\leq$~25$^{o}$ (left panel) and 25$^{o}\leq\theta_{lab}\leq$~50$^{o}$ (right panel). N-body Coulomb trajectory calculations 
\linebreak[4]
(lines in Fig. 24) indicated that the data are consistent with very short emission times, $\tau<$100 fm/c. The authors of Ref. \cite{ref93} have also observed that the study of kinetic energy gated IMF-IMF correlation functions is sensitive to the details of the fragment emission topology (surface/volume emission processes) in the initial state. However, no definitive quantitative results could be deduced because of the unknown sensitivity to other correlations and conservation laws in the initial state \cite{ref93}.

The perspective of extracting emission times from IMF-IMF correlation functions seems to be quite attractive in order to better understand even the evolution of emission mechanisms with the excitation energy deposited into the excited system undergoing the multifragment decay. The study of such evolution with the excitation energy has been performed both in central collisions at intermediate energies and in peripheral collisions at relativistic energies. In the next section we will present the results obtained from IMF emission-time measurements in central collisions where the excitation energy is controlled by the incident beam energy. In the following section we will show similar emission-time studies extended to the decay of excited target spectators after bombardment by relativistic-energy light probes. In this case, the decay of the system is mostly governed by the deposited excitation energy, without strong contributions from collective motion that represent a relevant phenomenon in the case of central collisions. These two sections will allow us to have an idea about how the emission times of IMFs are correlated with the degree of excitation of the decaying system.

 \subsection{Emission timescales in central collisions: sensitivity to the incident energy}
 
 Mean emission lifetimes for multifragment final states produced in Kr+Nb reactions at incident energies E/A=35, 45, 55, 65 and 75 MeV/nucleon were studied in Ref. \cite{ref97}. The width of the Coulomb dip in the IMF-IMF correlation function at small reduced velocities was observed to increase as the bombarding energy increases from E/A=35 to E/A=55 MeV. This result is consistent with a reduction of IMF emission times as the violence of the collision is increased. The measured IMF-IMF correlation functions were compared to classical three-body Coulomb trajectory calculations performed with the code MENEKA \cite{ref99} and the extracted emission times are represented in Fig.~25 as a function of the incident energy. Long emission times, $\tau\approx$~400 fm/c, are observed at lower incident energies, E/A=35 MeV. As the incident energy is increased, the emission of IMFs is observed to occur over shorter timescales, until a saturation at a value of $\tau\approx$~100 fm/c is observed at E/A=55 MeV \cite{ref99}. This result is consistent with an evolution of the fragment emission mechanism from long-lived sequential processes, at lower incident energies, to more a simultaneous scenario, at higher incident energies. The extracted minimum lifetime, $\tau\approx$~100 fm/c, is comparable to the timescales of growing density fluctuations for nuclear matter in the low-density instability regions of the nuclear phase diagram.
 
\begin{figure}
\centering
\resizebox{0.85\columnwidth}{!}{
  \includegraphics{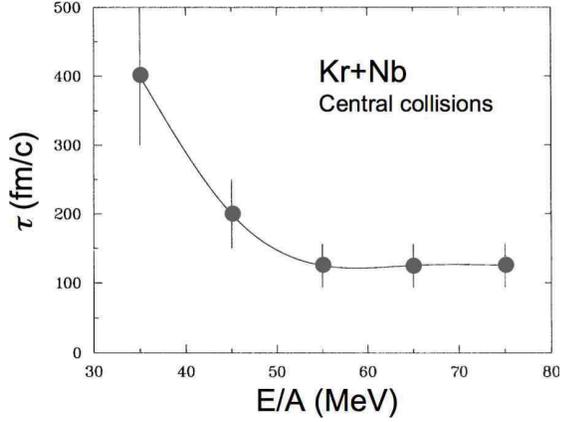}}
  \caption{IMF emission lifetimes extracted in the study of central Kr+Nb collisions at incident energies between E/A=35 and 75 MeV \cite{ref97}.}
\end{figure}
 
\begin{figure}
\centering
\resizebox{0.95\columnwidth}{!}{
  \includegraphics{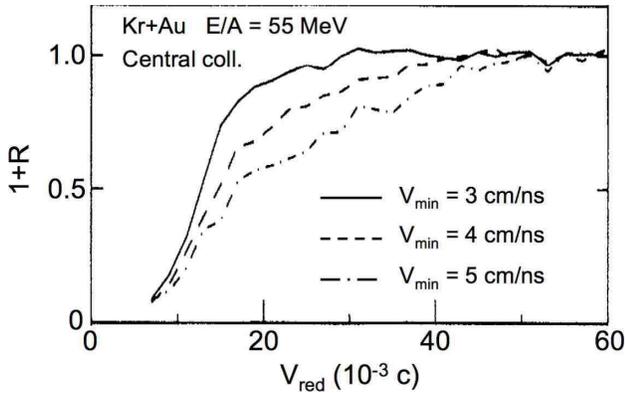}}
  \caption{IMF-IMF correlation functions measured in Kr+Au central collisions at E/A=55 MeV and corresponding to different gates on the velocity of the slower of the two coincident fragments \cite{ref98}.}
\end{figure}

The results shown in Fig. 25 are based on the assumption that the fragments are emitted from a system with a single freeze-out condition well localized in time. This scenario might be too simplified. Figure 26 shows IMF-IMF correlation functions measured in Kr+Au reactions at E/A=55 MeV for three different values of $v_{min}$, defined as the minimum velocity of the less energetic fragment of each pair. It is clearly seen that faster IMFs exhibit a wider Coulomb hole, indicating shorter emission time-scales. Similar results were observed also at E/A=35 and 70 MeV \cite{ref98}. The dependence of the emission lifetime on the velocity of the emitted fragments can be considered as an indication that a single freeze-out condition for fragment emission might not exist. This indication was confirmed by the calculations performed with the Expanding Evaporating Source model \cite{ref100} that predicted decreasing emission times for fragments with higher velocities \cite{ref98}. Models based on a simultaneous multifragmentation scenario were not capable of reproducing such evolutionary emission pattern. A similar evolutionary scenario in multifragment emission is also supported by the extraction of different emission times for different nuclear species \cite{ref101}. For instance, in the study of Kr+Au collisions at E/A=70 MeV, emission times of about 50~fm/c and 100~fm/c were found, respectively, for carbon and beryllium isotopes \cite{ref101}. 
 
The described results obtained in the study of central collisions have provided evidence for short emission times at higher energies. While this finding suggests a simultaneous scenario for multifragmentation, velocity-gated and Z-gated correlation studies suggest that evolutionary decay processes cannot be excluded \cite{ref98,ref100,ref101}. The definition of a freeze-out stage might be considered more involved than expected, requiring further experimental and theoretical investigations. In this respect, it would be important to compare experimental correlation functions to predictions of microscopic models that describe the whole dynamical evolution of multifragmenting nuclear systems. Similar studies have recently been performed \cite{ref96,ref102} by comparing IMF-IMF correlation functions for central Xe + Sn and Gd + U collisions at E/A=32 MeV, measured with the INDRA detector, to simulations performed with the BOB model (Brownian One Body dynamics) based on the BNV (Boltzmann-Nordheim-Vlasov) approach 
\cite{ref95,ref103,ref104}. The authors have studied higher order correlations by selecting different decay channels based on fragment charges and specific event topologies \cite{ref96,ref102}. The conclusions of these studies seems to support freeze-out times of the order of 200-240 fm/c and fragment spatial distributions of the order of 3-4~$V_{0}$ in Xe+Sn and 8~$V_{0}$ in Gd+U reactions, with $V_{0}$ being the volume of the source at normal density, $V_{0}=(4\pi/3)(1.2)^{3}A_{tot}$~fm$^3$.
 
The study of central collisions is generally complicated by the presence of collective motion. The explosive nature of such systems can decrease the effective IMF emission times. It is, therefore, not easy to correlate the IMF emission times shown in Fig. 25 with the deposited excitation energy. 
 
In the next subsection we will present measurements of IMF emission times in target spectators where estimates of the actual thermal deposited excitation energy were made. Such studies are less dependent on the dynamical effects induced by collective motion and can provide better links to the occurrence of a liquid-gas phase transition in nuclear matter driven by thermal effects.
 
\begin{figure}
\centering
\resizebox{0.95\columnwidth}{!}{
  \includegraphics{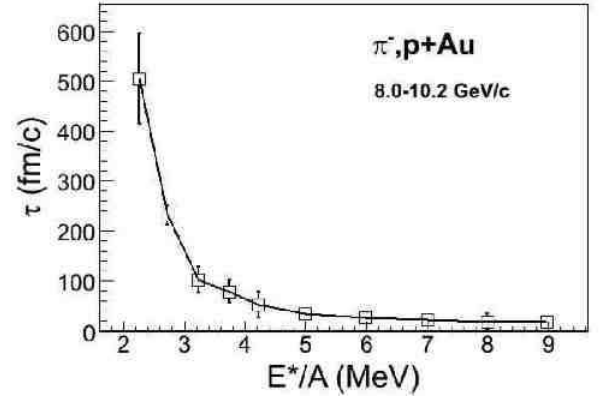}}
  \caption{IMF emission lifetimes as a function of the deposited excitation energy measured in target spectators decay induced by the bombardment $\pi^{-}$ and $p$ beams at incident momentum of 8.0 and 10.2 GeV/c \cite{ref94}.}
\end{figure}

\subsection{Evolution of emission times with excitation energy in target spectators}

Fragmentation phenomena of target-spectator fragmentation induced by light probes at relativistic energies are one of the best tools to investigate thermally driven phase transitions. In the study of $\pi^{-}$,$p$+Au at 8.0, 8.2, 9.2 and 10.2 GeV/c, IMF-IMF correlation functions from the decay of Au target spectators were studied \cite{ref94}. The evolution of IMF emission times with the excitation energy per nucleon is represented in Fig. 27. As the deposited excitation energy increases, emission lifetimes decrease from $\tau\approx$~500 fm/c at excitation energies E*/A~2.5 MeV to a saturating value of about $\tau\approx$~20-50 fm/c for excitation energies above 5 MeV/nucleon. These results indicate a transition from a surface evaporation-like emission at low excitation energies towards a bulk simultaneous multifragmentation scenario above excitation energies of the order of E*/A=5 MeV \cite{ref94}. Furthermore, the extracted emission times seem to be comparable with timescales of thermodynamical fluctuations leading to liquid-gas phase transitions in nuclear matter. The decreasing emission times should be related to the increasing thermal excitation energy deposited into the system. Contrary to the case of central collisions presented in the previous subsection, only a small collective-motion component caused by the thermal expansion of the system should exist and the dynamics of the decay is mostly dominated by thermodynamical aspects.

\section{Internal excitation energy of complex fragments}  

Light-particle-fragment correlation functions have been 
\linebreak[4]
used to perform a thermal characterization of the emitting source by providing experimental information on the size and the internal excitation energy of the primary fragments \cite{ref105,ref106,ref107,ref108,ref109}. 

Several statistical and dynamical models predict very different internal excitation energies of the primary fragments. The quantum molecular dynamic (QMD) model \cite{ref4,ref110,ref111,ref112} and the microcanonical metropolis Monte Carlo (MMMC) model \cite{ref113} predict a rather low excitation energy for primary fragments. Antisymmetrized molecular dynamics \cite{ref114,ref115,ref116} (AMD), stochastic mean field simulations \cite{ref95,ref103,ref104,ref117}, the statistical multifragmentation model \cite{ref118,ref119} (SMM) and microcanonical multifragmentation models \cite{ref120,ref121,ref122} predict moderately hot primary fragments. Moreover, the mechanism responsible for the formation of fragments differs from a model to another. In dynamical models, the formation of the fragments and their excitation energy depend not only on the collision dynamics but also on the procedure employed for performing the cluster recognition \cite{ref123,ref124,ref125}. On the other hand, in the case of statistical models, the excitation energy of the fragments is an assumption of the model itself used to calculate the statistical weights of the partitions. Therefore, experimental estimates of the importance of secondary decays and of the size of the primary fragments can provide very significant constraints and crucial tests for the theoretical models. A direct measurement of these quantities would be also interesting for various aspects: the extraction of certain physical information, such as the rate of statistically to dynamically emitted particles or the caloric heat capacity \cite{ref14,ref15}, would be less dependent on model assumptions.

Previous studies have shown that it is possible to extract the intrinsic properties of the fragments independently of the mechanism of their formation. These studies are based on the measurement of relative-velocity correlation functions between fragments and light charged particles (LCP).  The aim of this section is to give a review of the work dedicated to the determination of the sizes and excitation energies of primary fragments. An excitation function of these quantities will be given for Xe + Sn system for the incident energy range E/A = 30-50 MeV \cite{ref105,ref107,ref108,ref109}. We will present results on the study of central Kr+Nb collisions at E/A = 45 MeV \cite{ref106} and of quasi projectiles produced in peripheral Xe+Sn collisions at E/A = 100 MeV \cite{ref126}. These studies were conducted by using either the Washington University Dwarf Ball/Wall array \cite{ref127} or the 4$\pi$ INDRA multidetector \cite{ref63,ref64,ref65}. It is important to note that the same data can provide information on the space-time extend of the outgoing fragments, but in this section we will concentrate only in the measurement of their excitation energy.

\subsection{Reconstruction of the primary fragments}

At intermediate energies, there are at least two extreme mechanisms for the production of light particles, (i) in nucleon-nucleon collisions and (ii) by statistical evaporation from excited sources. The former occurs during the first stage of the collision when the dynamics play an important role. This process is called direct or pre-equilibrium emission. The second process is supposed to occur at later stages of the reaction as a secondary emission from excited primary fragments. Between these two stages, continuous emission may occur from other sources that are difficult to define. In multifragmentation events, it is possible to isolate the secondary statistical component if the fragments formed are not too excited, so that the time scale associated with their decay is much larger than the time scale of their production. 

\begin{figure}
\centering
\resizebox{0.85\columnwidth}{!}{
  \includegraphics{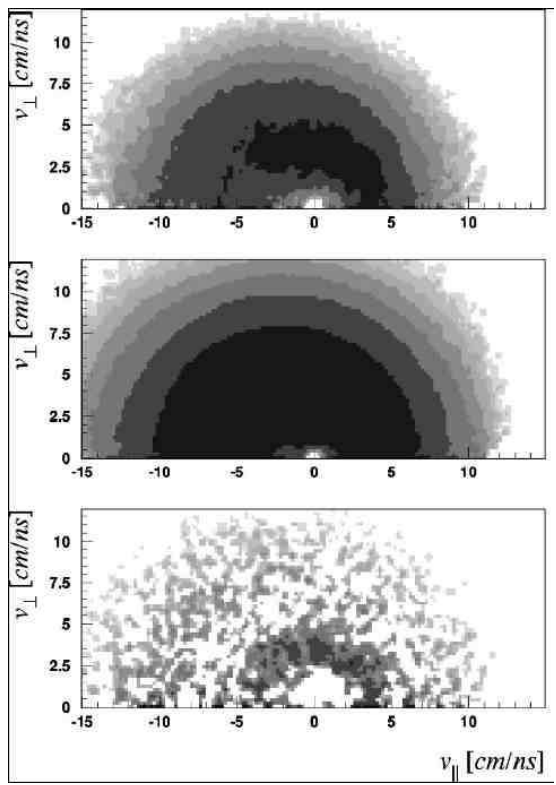}}
  \caption{Reaction system: Kr + Nb at E/A = 45 MeV. Parallel velocity-transverse velocity invariant diagram for protons in the center of mass of the heaviest fragment detected in coincidence in each event. Top panel: correlated events; Middle panel:  uncorrelated events; Bottom panel: difference between correlated and uncorrelated events \cite{ref106}.}
\end{figure}

\begin{figure}
\centering
\resizebox{0.85\columnwidth}{!}{
  \includegraphics{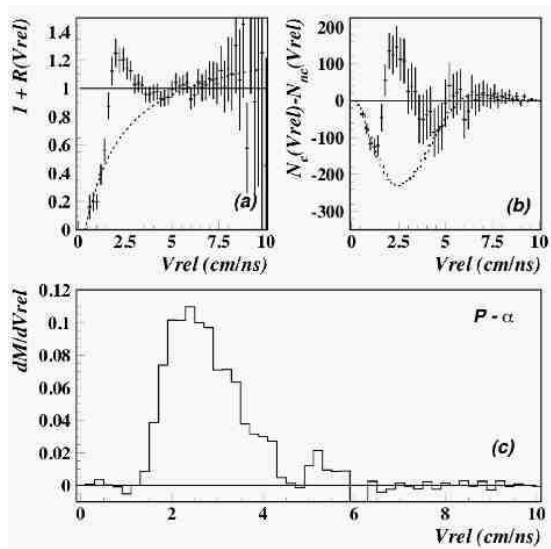}}
  \caption{(a) Phosphorus-alpha correlation measured in central collisions of Xe + Sn at E/A = 32 MeV. (b) Difference function. (c) Velocity spectrum of $\alpha$ particles in the center of mass of the phosphorus fragments, obtained by subtracting the background (dashed line in (b)) from the the difference function (data points in (b)), from Ref. \cite{ref109}.}
\end{figure}

In this context the primary fragment excitation energy can be deduced from the multiplicity of its associated evaporated LCPs. Figure 28, taken from Ref. \cite{ref106}, shows the projection of the relative velocity between the heaviest fragment and the protons detected in coincidence and in the case of Kr+Nb collisions at E/A = 45 MeV. The components V$_{||}$ and V$_{\perp}$ are the projections of the relative velocity onto the axis representing the fragment direction in the center of mass frame and into a plane perpendicular to that axis. The top panel corresponds to raw coincidence data, the middle panel corresponds to the background distribution and the bottom panel displays the result of a subtraction of this background from the total LCP emission. This bottom panel clearly shows a ring surrounding the IMF location in velocity space (V$_{||}$ = V$_{\perp}$= 0). This feature may correspond to the Coulomb ring associated with the proton emission from the heaviest fragment. In order to estimate quantitatively the amount of evaporated protons from secondary decay processes, one needs to estimate the contributions induced by other background effects that have to be properly subtracted. Few methods have been employed to perform such background estimation and subtraction \cite{ref105,ref106,ref107,ref108,ref109} and they are all based on a scenario deduced from Boltzmann-Nordheim-Vlasov \cite{ref128} calculations. Multifragmentation is described as a two-step process. The first step is the cooling of the initial fused system through a sequence of light-particle-emission processes. The second step is the fragmentation of the smaller remaining source where the remaining excitation energy is shared between a fixed number of primary fragments. These fragments then decay sequentially while moving apart under the influence of the Coulomb force. An initial radial velocity can be added to the Coulomb motion in order to mimic a possible expansion of the source. Based on such a scenario and on model predictions, two approaches have been suggested to estimate the background of the correlation function, differing in the way one compares calculations to experimental data. In one case, the shape of the background is deduced, normalized to the data and then subtracted \cite{ref105,ref107,ref108,ref109}. In the other case, the input parameters of the calculations are adjusted until the data points are reproduced (kind of back-tracing) \cite{ref106}. After the background is subtracted, one can access the amount of evaporated particles from the primary fragments. In the following subsection we will describe how the technique is applied to experimental data and what physical information has been extracted.

\subsection{Application to data and experimental results}

Figure 29 shows the experimental relative velocity correlation function (a) and the difference function (b) of the phosphorus - alpha pairs measured in the central collisions Xe + Sn at E/A = 32 MeV. To build the correlation function shown in this figure, Eq. (2) has been used by replacing the momenta $\vec{p}_{1}$  and $\vec{p}_{2}$ with the velocities $\vec{v}_{1}$ and $\vec{v}_{2}$ of a given fragment (here is a phosphorus) and the LCP (here an $\alpha$ particle). The uncorrelated events have been constructed by the event-mixing technique : For a given fragment in an event $i$ having $n$ LCP, we take randomly n LCP from n different events, then the relative velocity of this fake event is calculated. The dotted lines represent the background \cite{ref105,ref106,ref107,ref108,ref109}. This background should contain any other contribution, which is not evaporated from parents of phosphorus. The relative velocity distribution presented in Fig. 29c, obtained by subtracting the background from the difference function, can be considered as the deduced contribution of $\alpha$ particles evaporated by parents of phosphorus. 

\begin{figure}
\centering
\resizebox{0.9\columnwidth}{!}{
  \includegraphics{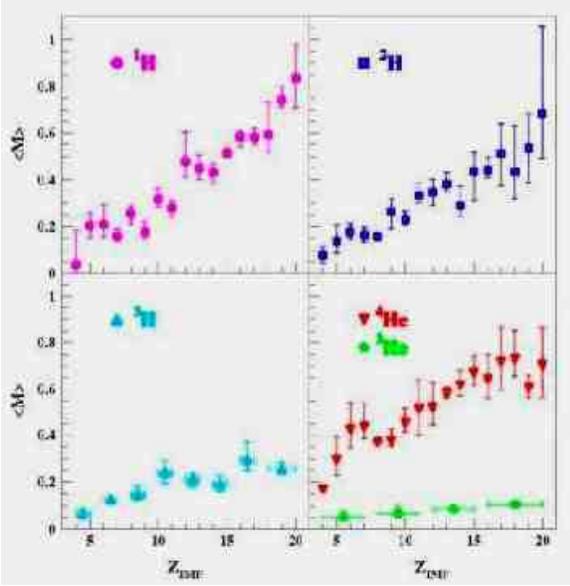}}
  \caption{Average secondary multiplicities per fragment of the evaporated  hydrogen and helium isotopes as a function of the atomic number of the detected fragments for Xe+Sn central collisions at E/A = 50 MeV \cite{ref105}.}
\end{figure}

This procedure was applied to all fragment-LCP pairs produced in the central collisions Xe + Sn at E/A = 32, 39, 45 and 50 MeV. Thus, for each detected fragment, the average multiplicities of the light particles evaporated by the primary fragment was determined. These multiplicities increase slightly with the size of the fragment. However they remain weak, not exceeding the value of 1.5 at all incident energies, implying that the excitation energy of the primary fragments is particularly moderate. An example of the extracted multiplicity is given in Fig. 30 for central collision of Xe + Sn at E/A = 50 MeV. The multiplicity of a given LCP does not change much with incident energy. From the spectra of the evaporated light charged particles, the average kinetic energy has been extracted. 

\begin{figure}
\centering
\resizebox{0.98\columnwidth}{!}{
  \includegraphics{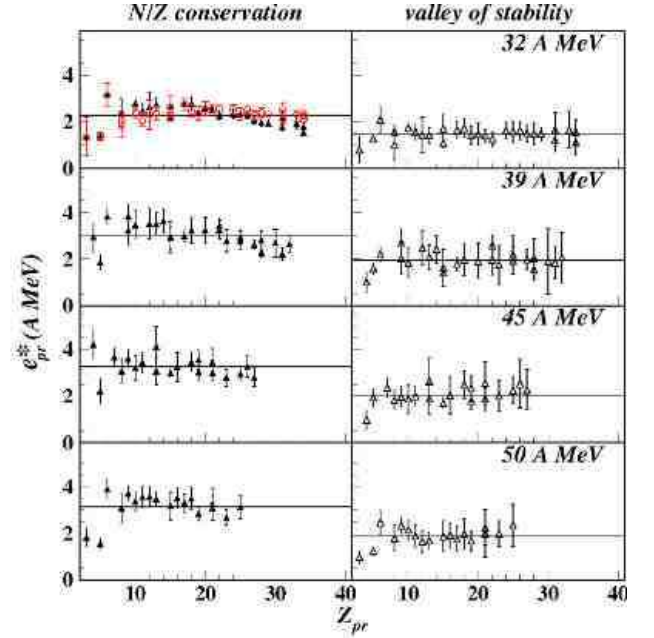}}
  \caption{Average excitation energy per nucleon of the primary fragments as a function of their charge for central Xe+Sn collisions at E/A = 32, 39, 45 and 50 MeV. The horizontal lines represent a fit to the average excitation energies with a Z-independent value for each bombarding energy. Left panels: the primary fragments have the same N/Z as the combined system. Right panels:  the fragments are produced in the valley of stability. The masses of detected fragments are assumed to follow the valley of stability except for the data points represented by open circles where the EAL assumption is used \cite{ref109}.}
\end{figure}

Both observables (average multiplicity and kinetic energy) have been used in order to deduce the average charge of the primary fragments, $<Z_{pr}>$, their average mass, $<A_{pr}>$ and their excitation energy, $E^{*}$. The average charge of the primary fragment is calculated as the sum of the charge of the detected fragment, $Z$ and the charge of all light charged particles correlated to that fragment, $<Z_{LCP}>$. This last quantity is defined by $<Z_{LCP}>=\sum_{i}z_{i}\cdot<M_{i}>$ , where  $z_{i}$ and $<M_{i}>$ are the charge and the average multiplicity of the evaporated particles, $i=p, d, t, ^{3}He, \alpha$. Since the isotopes of the detected fragments are not resolved and the neutrons are not detected, two extreme assumptions were necessary in order to determine the primary mass of the fragment. The first assumption consists of assuming that the detected fragment is produced in the valley of stability. The second assumption considers two cases: i) the primary fragments are produced in the valley of stability; ii) the primary fragments keep the same ratio N/Z as the initial system. The obtained charge of the primary fragments varies between 1 and 5 units of charge in addition to the charge of the detected fragment. The mass of the primary fragments depends also on the considered assumptions. The average multiplicity of the neutrons is deduced by the mass conservation, knowing the mass of the primary fragment, of the detected fragment and that of the secondary light particles. Multiplicities of about 7 are reached, but depend strongly on the assumption made on the mass of the primary fragment. 

\begin{figure}
\centering
\resizebox{0.8\columnwidth}{!}{
  \includegraphics{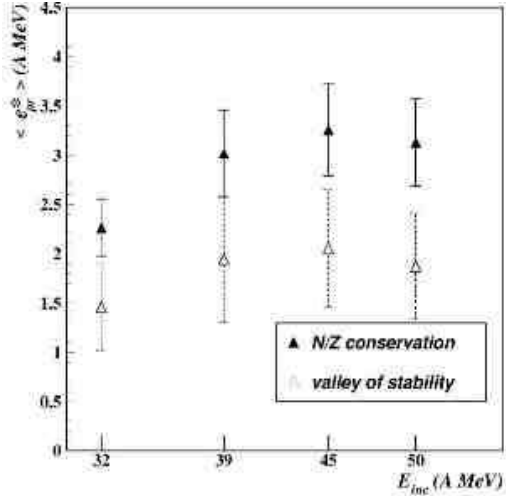}}
  \caption{Average excitation energy per nucleon of primary fragments produced in central collisions of the Xe+Sn system as a function of incident energy. The black and open symbols correspond to two hypotheses of the fragment masses. The vertical lines represent the standard deviation from the mean values \cite{ref109}. }
\end{figure} 

Finally with the help of all these variables, it was possible to apply a calorimetry method to determine the excitation energy of the primary fragments $E_{pr}^{*}$. Figure 31 shows the average excitation energy per nucleon $E_{pr}^{*}/A$  of the primary fragment as a function of its detected charge, and this for four incident energies and the two assumptions on the mass of the primary fragment. The horizontal line in Fig. 31 represents the average value $<e_{pr}^{*}>$ within the range of all studied primary fragments.  Besides nuclei with low charges, all the experimental points lie on this straight horizontal line within error bars. In other words, whatever the incident energy and the mass assumption considered, the excitation energies per nucleon of the primary fragments are constant. Figure 32 shows the evolution of the average value with the bombardment energy. The vertical bars represent the standard deviations from the mean values. They are small and do not exceed 1 AMeV, consolidating the constancy of the value $<e_{pr}^{*}>$. In the case of the N/Z conservation assumption, the excitation energy per nucleon increases from 2.2 AMeV at E/A = 32 MeV up to a saturation value of 3 AMeV at E/A $\geq$~39 MeV. In the case of the other assumption, where the primary fragments are produced on the valley of stability, the values saturate also but at lower energy. The constancy observed in Fig. 32 of the excitation energy per nucleon for the various primary fragments, can suggest that thermodynamical equilibrium was reached during the disintegration of the system. On the other hand, the saturation of $<e_{pr}^{*}>$ can indicate that the fragments have reached their limiting excitation energy per nucleon (or their limiting temperature) \cite{ref130,ref131}. 

\begin{figure}
\centering
\resizebox{0.98\columnwidth}{!}{
  \includegraphics{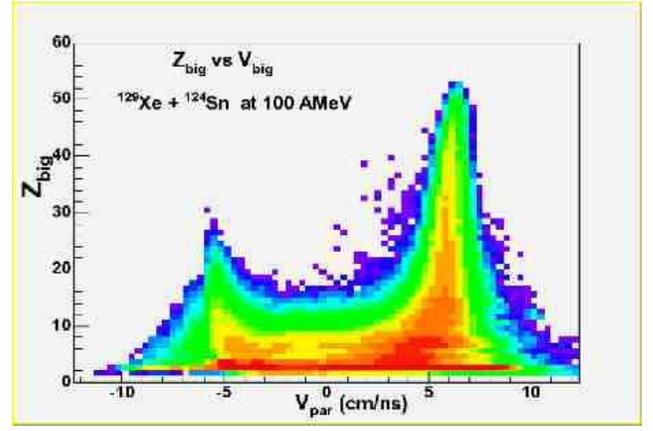}}
  \caption{ Atomic number of the heaviest fragment in the event as a function of center of mass parallel velocity, for Xe+Sn at E/A = 100 MeV reaction.}
\end{figure} 
 
\begin{figure}[hb]
\centering
\resizebox{0.98\columnwidth}{!}{
  \includegraphics{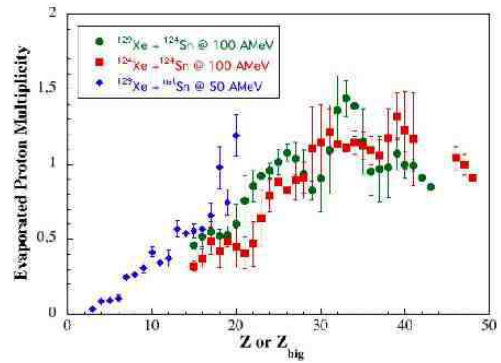}}
  \caption{Evaporated proton multiplicity as function of the atomic number of the emitter: quasi-projectile or fragments formed in central collisions. The systems presented here are indicated in the figure.}
\end{figure} 

\begin{figure*}
\centering
\resizebox{1.6\columnwidth}{!}{
  \includegraphics{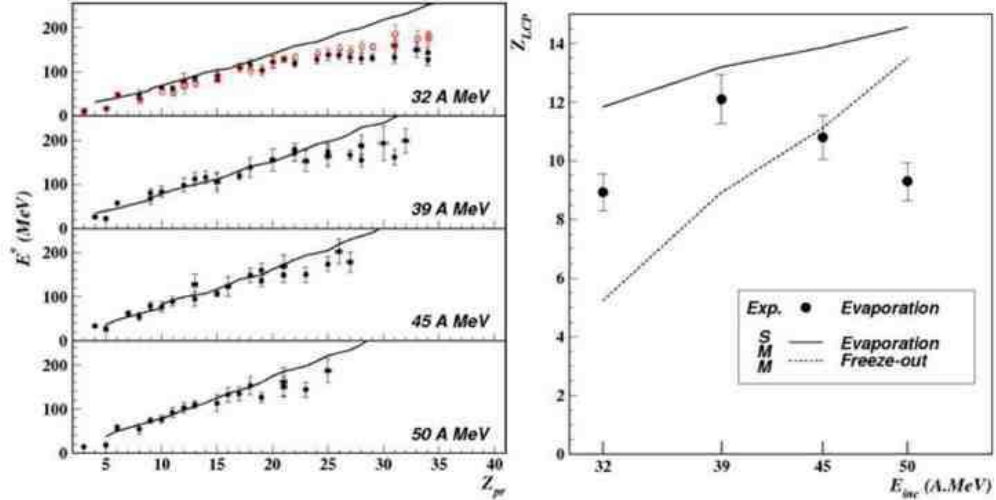}}
  \caption{Comparison between SMM calculations (curves) and data (symbols) are presented. Left panel: Average excitation energy of the primary fragments as a function of their atomic number for central Xe+Sn collisions at E/A= 32-50 MeV. Right panel: Total charge contributions of secondary evaporated particles $Z_{LCP}$. The dotted histogram represents the calculated freeze-out contribution \cite{ref109}.  }
\end{figure*} 

How to choose between the two assumptions of mass on the primary fragments? This choice was dictated by calculations with a statistical code GEMINI \cite{ref132}. This code is very well suited at low excitation energies not exceeding 3-4 AMeV. The procedure consists in using the deduced primary-fragment characteristics, $Z_{pr}$, $A_{pr}$ and $E_{pr}^{*}$ as input parameter, letting them decay and comparing the resulting multiplicities of the evaporated particles to the experimental values. It was shown in Ref. \cite{ref105} that the assumption of the N/Z conservation was reasonable.

The average multiplicities of secondary light charged particles allowed an intrinsic characterization of the primary fragments, but it can also be used to give valuable information on multifragmentation events. Indeed, the ratio of the secondary evaporated LCP multiplicities to the total detected LCP multiplicities gives the fraction of thermally 
produced particles. It has been shown in Refs. \cite{ref108,ref109} that the maximum proportion of evaporated particles does not exceed, on average, 35\% of the total number of produced light charged particles. For incident energies between E/A = 32 and 39 MeV, the proportion of evaporated particles increases, thus reflecting the increase of the excitation energy of the fragments observed in Fig. 32. Beyond E/A=39 MeV, this proportion decreases to reach 23\% at E/A=50 MeV while the excitation energy of the primary fragments remains constant. It should be noted that the extracted proportion of secondarily evaporated particles constitutes a lower limit, because it does not contain 
the contributions which can come from the disintegration of the unstable nuclei like the $^{8}$Be, $^{5}$Li etc. and the decay of the short-lived excited states \cite{ref133}. Similar results have been extracted also from the study of Kr+Nb reactions at E/A = 45 MeV \cite{ref106}. It has been shown that i) the excitation energy of the primary fragments does not exceed 2.5 AMeV, ii) about 
80\% of the detected LCPs do not originate from the secondary statistical decay of the primary fragments. 

\subsection{Comparison to models}

The described techniques have been applied to the quasi-projectile (QP) nuclei formed in Xe + Sn collisions at E/A = 100 MeV in order to extract their thermal characteristics \cite{ref126}. Figure 33 gives an overview of the collision.  It shows the atomic number of the heaviest fragments in the event as a function of its velocity parallel to the beam direction. One can observe clearly two main components, one centered at projectile velocity and the other one centered at target velocity (the heavy quasi target fragments are not detected since their velocity do not exceed the energy threshold of the used detectors). Between the two components one can also observe an emission of fragments at mid-rapidity and characterized by a low charge and a high cross section. 

This preliminary result is surprising since the excitation energies of the spectator at E/A=100 MeV and  participant fragments at E/A=50 MeV are almost the same. Is it an indication of thermal energy saturation? Or are we dealing with the same fragment production mechanism? The question is open.

The correlation function method has been applied to extract the secondary evaporated light charged particles from each QP. Figure 34 shows results for two systems having different isospin: $^{129}$Xe + $^{124}$Sn, $^{124}$Xe+$^{124}$Sn. The figure represents the extracted proton multiplicity as a function of the charge of the QP emitter. The data points obtained above for central collisions and same system but at a lower beam energy E/A=50 MeV are superimposed on the same figure. All data points follow the same systematic: the evaporated proton multiplicity increases with the charge of the quasi-projectile or with the charge of the fragment (in the case of central collisions). The proton multiplicity values do not exceed 1.5 and are compatible with an excitation energy of 2-3 AMeV.  

The experimental estimate of the secondary decay component can be compared
with the predictions of statistical multifragmentation models such as the SMM \cite{ref118,ref119} 
or the micro-canonical model of Raduta et al. \cite{ref120,ref121,ref122} which explicitly consider fragment excitation. The comparison of the extracted quantities with the calculations can indeed constitute a crucial test of the basic assumptions of these models. In the MMMC \cite{ref113} approach, the 
excitation of primary fragments is implicitly included in the neutron production at freeze-out, 
and a direct comparison 
is not possible. 

\begin{figure}
\centering
\resizebox{0.9\columnwidth}{!}{
  \includegraphics{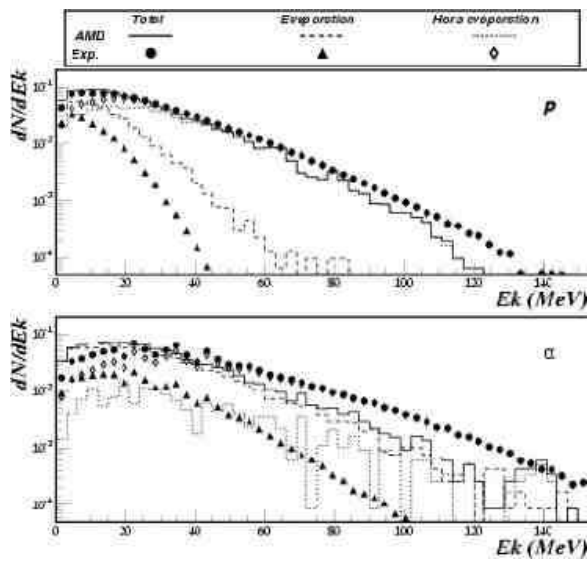}}
  \caption{Proton and alpha energy spectra produced in central collisions of Xe+Sn at E/A=50 MeV. The symbols are the data and curves represent AMD calculations. The dots (continuous histogram) represent the total energy spectra, the triangles (dotted histogram) represent the secondary evaporation contribution and the open symbols (hatched histogram) represent the direct early emission. }
\end{figure} 

Statistical Multifragmentation model (SMM) calculations have been performed in order to reproduce the thermal component. The version of SMM used in \cite{ref108,ref109} gives access to the freeze-out configuration, i.e., to primary-fragment characteristics before secondary decay. 
Figure 35 shows the results of this calculation. The left panel in this figure represents the excitation energy in MeV of the primary fragments as a function of their atomic number (same data points as in Fig. 31 with the only difference being the use of MeV/nucleon as units of excitation energies). The right panel represents the total charge contributions of secondary evaporated particles, $Z_{LCP}$. The data indicate that a maximum LCP evaporation is obtained for E/A = 39 MeV and a decrease of thermal contribution takes place above this energy. Although the internal excitation energy of the primary fragments is well reproduced at each incident energy, the trend of the total charge of secondary LCPs, $Z_{LCP}$, with the beam energy is not reproduced. This discrepancy can be understood if we consider the increasingly important effects of the collision dynamics as the beam energy increases. Direct emissions of LCP increase with increasing incident energy, while the proportion of the thermal contribution decreases.

The obtained results have also been compared to dynamical calculations. In order to understand the mechanism responsible for the saturation of the excitation energy of the primary fragments observed in Fig. 33, AMD calculations \cite{ref114,ref115,ref116} for central collisions of Xe + Sn at E/A=50 MeV have been performed.
A reasonably good agreement for the charge distribution, the charge of the heaviest fragment in the event and average kinetic energy of the fragments has been obtained \cite{ref108,ref126,ref134}. The comparison of such calculations to the experimental data is shown in Fig. 36. It represents the energy spectra of proton and alpha particles produced as free particles in AMD (direct emission) and secondary particles evaporated from the calculated decay of the excited primary fragments. The two contributions are compared to experimentally deduced secondary particles with the method described above. AMD predicts reasonably well the total energy spectra of protons  (78\% 
of total emission), but it fails to reproduce the thermal contribution (22\%).

\section{Conclusions}

The study of heavy-ion collisions at intermediate energies has provided important information about the mechanisms of multifragmentation and their links to a possible nuclear liquid-gas phase transition. Our capability of fully understanding these phenomena depends strongly on how well one can identify and characterize the thermal and dynamical properties of fragment and particle emitting sources. In this chapter we have presented a review of the most significant results that different research groups have achieved in the last decades by using two-particle correlation techniques. 

Intensity interferometry techniques have 
demonstrated that we have {\it space-time probes capable of measuring sizes as small as 10$^{-15}$m and time intervals as short as several 10$^{-23}$sec}. This can be viewed as an important advance in the field of heavy-ion physics. These space-time characterization techniques have provided valuable information about the size/volume, the density of decaying nuclear systems, as well as quantitative estimates of emission lifetimes and particle chronology. The extracted results 
demonstrate that the emitting sources produced in heavy-ion collisions 
form a quite complex system: different particles are emitted at different times and by different mechanisms.   
In this respect, it seems clear that a complete space-time characterization of nuclear reactions requires a study of multiple correlations extended to all particle species and including neutrons. Proton-neutron chronology is also indicated as an important candidate to explore the density dependence of the symmetry energy which represents one of the most challenging perspectives for the future. 

The recent introduction of imaging techniques has certainly changed our interpretation of correlation observables. The source sizes extracted with this technique can differ significantly from those extracted from other approaches. This difference suggests that a detailed shape analysis of correlation observables is necessary in order to obtain a correct determination of nuclear densities and lifetimes. 
In this chapter we have also shown how 
techniques were found to quantify the 
strength of secondary decays. Light particle-fragment correlations have been used to extract information about the excitation energy of primary unstable fragments. The possibility of constraining secondary decays represents an important perspective: several observables are indeed affected by the unknown contributions to the experimental spectra from different emission mechanisms dominating different stages of the reaction.

Despite the large amount of information that the scientific community has been able to extract, certain aspects have not been fully explored and require further research. For instance, from an experimental point of view, higher isotopic resolution could provide an important opportunity to investigate isotopically 
resolved fragment-fragment correlation functions. This perspective is very important in view of clarifying the indications of evolutionary freeze-out conditions in heavy-ion collisions and their possible links to the equation of state of asymmetric nuclear matter, relevant to astrophysical environments like neutron stars and supernova explosions. Higher angular resolution and a large solid angle coverage are important requirements in order to increase the quality of measured correlation observables and to explore their features while having a complete characterization of the collision event (impact parameter, reaction plane, exact determination of velocity vectors, etc.). 
Higher resolution devices will be the key to the perspective of extending imaging techniques to several particle species, improving complex particle correlation analyses that provide information about the space-time properties at freeze-out and about the excitation energies of primary fragments. Other fields of nuclear physics research will certainly profit from the existence of detector setups that are characterized by outstanding correlation capabilities. This is particularly the case for the nuclear structure groups working on spectroscopic properties of exotic nuclear systems explored with the future radioactive-ion-beam accelerator facilities.

The close interaction between the theoretical and experimental communities will certainly contribute significantly in improving our capability of characterizing emitting sources. The use of more powerful experimental  setups and the implementation of full quantum multi-body approaches to correlation functions promise to provide unambiguous information about the thermal and dynamical scenarios characterizing multifragmention phenomena and their links to the equation of state of nuclear matter.

\newpage 

\begin{thebibliography}{9}

\bibitem{ref1}
W.G. Lynch, Annu. Rev. Nucl. Part. Sci 37,
  493 (1987).

\bibitem{ref2}
L.G. Moretto, and G.J . Wozniak, Annu. Rev. Nucl. Part. Sci. 43, 379 (1993).

\bibitem{ref3}
Proceed. Int. Workshop XXVII on Gross Properties of Nuclei and Nuclear Excitations,  1999, edit. H. Feldmeier et al., GSI, Darmstadt.

\bibitem{ref4}
J. Aichelin, Phys. Rep. 202, 233 (1991).

\bibitem{ref5}
C.P. Montoya, et al., Phys. Rev. Lett. 73, 3070 (1994). 

\bibitem{ref6}
Y. Larochelle et al., Phys. Rev. C 59, R565 (1999).
 
\bibitem{ref7}
E. Plagnol et al., Phys. Rev. C 61, 014606 (1999).
 
\bibitem{ref8}
J. Lukasik et al., Phys. Lett. B, 566, 76 (2003). 

\bibitem{ref9}
T. Lefort et al., Nucl. Phys. A 662, 397 (2000). 

\bibitem{ref10}
Ph. Eudes, Z. Basrak and F. Sebille, Phys. Rev. C 56, 2003 (1997). 

\bibitem{ref11}
J. Pochodzalla et al., Phys. Rev. Lett. 75, 1040 (1995).

\bibitem{ref12}
A. Chbihi et al., Eur. Phys. J. A5, 251 (1999).

\bibitem{ref13}
F. Gulminelli and Ph. Chomaz, Phys. Rev. Lett. 82, 1402 (1999).

\bibitem{ref14}
Ph. Chomaz, V. Duflot and F. Gulminelli, Phys. Rev. Lett. 85, 3587 (2000).

\bibitem{ref15}
M. D'Agostino et al., Nucl. Phys. A699, 795 (2002).

\bibitem{ref16}
W. Bauer, C.K. Gelbke and S. Pratt, Annu. Rev. Nucl. Part. Sci. 42, 77 (1992).

\bibitem{ref17}
D.H. Boal, C.K. Gelbke, B.K. Jennings, Rev. Mod. Phys. 62, 553 (1990).

\bibitem{ref18}
U. Heinz and B.V. Jacak, Annu. Rev. Nucl. Part. Sci. 49, 529 (1999).

\bibitem{ref19}
R. Hanbury Brown and R.Q. Twiss, Phil. Mag. 45, 663 (1954); Nature 177, 27 (1956).

\bibitem{ref20}
R. Hanbury-Brown and R.Q. Twiss, Nature 178, 1046 (1956).

\bibitem{ref21}
G. Goldhaber et al., Phys. Rev. Lett. 3, 181 (1959); Phys. Rev. 120, 300 (1960).

\bibitem{ref22}
M. Lisa, S. Pratt, R. Soltz and U. Wiedemann, Annu. Rev. Nucl. Part. Sci. 55, 357 (2005).

\bibitem{ref23}
S.E. Koonin, Phys. Lett. B 70, 43 (1977).

\bibitem{ref24}
S. Pratt and M.B. Tsang, Phys. Rev. C 36, 2390, (1987).

\bibitem{ref25}
M.A. Lisa et al., Phys. Rev. Lett. 70, 2545 (1993).

\bibitem{ref26}
R. Kotte et al., Eur. Phys. J. A 6, 185 (1999).

\bibitem{ref37}
S. Fritz et al., Phys. Lett. B461, 315 (1999).

\bibitem{ref25a}
R. Kotte et al., Phys. Rev. C51, 2686 (1995)

\bibitem{ref27}
W.G. Gong et al., Phys. Rev. C43, 1804 (1991).

\bibitem{ref28}
W.G. Gong et al., Phys. Rev. Lett. 65, 2114 (1990).

\bibitem{ref29}
B. Davin et al., Nucl. Instr. and Meth. A473, 302 (2001).

\bibitem{ref30}
T.C. Awes et al., Phys. Rev. Lett. 61, 2665 (1988).

\bibitem{ref31}
D.O. Handzy et al., Phys. Rev. C 50, 858 (1994).

\bibitem{ref32}
M.A. Lisa et al., Phys. Rev. C 49, 2788 (1994).

\bibitem{ref33}
P.A. DeYoung et al., Phys. Rev. C 39, 128 (1989).

\bibitem{ref34}
R. Kotte et al., Eur. Phys. J. A23, 271 (2005).

\bibitem{ref35}
C. Schwarz et al., Nucl. Phys. A 681, 279 (2001).

\bibitem{ref36}
V. Serfling et al., Phys. Rev. Lett. 80, 3928 (1998).

\bibitem{ref38}
W.G. Gong, W. Bauer, C.K. Gelbke and S. Pratt, Phys. Rev. C43, 781 (1991).

\bibitem{ref39}
F. Zhu et al., Phys. Rev. C44, R582 (1991).

\bibitem{ref40}
R. Ghetti et al., Nucl. Phys. A674, 277 (2000).

\bibitem{ref41}
G. Verde et al., Phys. Rev. C65, 054609 (2002).

\bibitem{ref42}
D.A. Brown and P. Danielewicz, Phys. Lett. B398, 252 (1997).

\bibitem{ref43}
D.A. Brown and P. Danielewicz, Phys. Rev. C57, 2474 (1998).

\bibitem{ref44}
D.A. Brown and P. Danielewicz, Phys. Lett. C64, 014902 (2001).

\bibitem{ref45}
R. Ghetti et al., Phys. Rev. C64, 017602 (2001).

\bibitem{ref46}
R. Ghetti et al., Phys. Rev. C62, 037603 (2000).

\bibitem{ref47}
B. Jakobsson et al., Phys. Rev. C44, R1238 (1991).

\bibitem{ref48}
N. Colonna et al., Nucl. Instr. and Meth. A421, 542 (1999).

\bibitem{ref40a}
W. D\"unnweber et al., Phys. Rev. Lett. 65, 297 (1990)

\bibitem{ref49}
M. Marques et al., Phys. Rev. Lett. 73, 34 (1994).

\bibitem{ref50}
A. Badala et al., Phys. Rev. Lett. 74, 4779 (1995).

\bibitem{ref51}
P. Sapienza et al., Phys. Rev. Lett. 73, 1769 (1994).

\bibitem{ref52}
Z. Chen et al., Phys. Rev. C36, 2297 (1987).

\bibitem{ref53}
D.H. Boal and J.C. Shillcock, Phys. Rev. C33, 549 (1986).

\bibitem{ref54}
J. Pochodzalla et al., Phys. Lett. 174B, 36 (1986).

\bibitem{ref55}
C.J. Gelderloos et al., Phys. Rev. C52, R2834 (1995).

\bibitem{ref56}
J. Pochodzalla et al., Phys. Rev. C35, 1695 (1987).

\bibitem{ref57}
C.B. Chitwood et al., Phys. Rev. Lett. 54, 302 (1985).

\bibitem{ref58}
J. Pochodzalla et al., Phys. Lett. B174, 36 (1986).

\bibitem{ref59}
Z. Chen et al., Phys. Lett. B199, 171 (1987).

\bibitem{ref60}
D. Fox et al., Phys. Rev. C38, 146 (1988).

\bibitem{ref61}
D.A. Cebra et al., Phys. Lett. B227, 336 (1989).

\bibitem{ref62}
D. Gourio et al., Eur. Phys. J. A7, 245 (2000).

\bibitem{ref63}
J. Pouthas et al. (INDRA collaboration), 
Nucl. Inst. and Meth. A357 (1995) 418.

\bibitem{ref64}
J.C. Steckmeyer et al., Nucl. Instr. and Meth. in Phys. Res. A361, 472 (1995).

\bibitem{ref65}
J. Pouthas et al. (INDRA collaboration), 
Nucl. Inst . and Meth. A369, 222 (1996).

\bibitem{ref66}
R. Lednicky and V.L. Lyuboshitz, Sov. Jour. Nucl. Phys. 35, 770 (1982).

\bibitem{ref67}
J. Pochodzalla et al., Phys. Lett. B 175, 275 (1986).

\bibitem{ref68}
D. Ardouin, Int. J. Mod. Phys. E6, 391 (1997). 

\bibitem{ref69}
C.J. Gelderloos and J.M. Alexander, Nucl. Instrum. Methods A 349, 618 (1994). 

\bibitem{ref70}
C.J. Gelderloos et al., Phys. Rev. Lett. 75, 3082 (1995). 

\bibitem{ref71}
C.J. Gelderloos et al., Phys. Rev. C 52, R2834 (1995). 

\bibitem{ref72}
R. Lednicky, V.L. Lyuboshitz, B. Erazmus, D. Nouais, Phys. Lett. B 373, 30 (1996). 

\bibitem{ref73}
S. Voloshin, R. Lednicky, S. Panitkin, and N. Xu, Phys. Rev. Lett. 79, 4766 (1997). 

\bibitem{ref74}
D. Ardouin et al., Phys. Lett. B 446, 191 (1999). 

\bibitem{ref75}
R. Ghetti et al., Phys. Rev. Lett. 87, 102701 (2001). 

\bibitem{ref76}
M.M. de Moura et al., Nucl. Phys. A 696, 64 (2001). 

\bibitem{ref77}
R. Ghetti et al., Phys. Rev. Lett. 91, 092701 (2003).

\bibitem{ref78}
R. Kotte et al., Eur. Phys. J. A6, 185 (1999). 

\bibitem{ref79}
S. Soff et al., Journ. Phys. G23, 789 (1997). 

\bibitem{ref80}
M. Di Toro, S.J. Yennello, Bao-An Li, contribution to this volume.   

\bibitem{ref81}
L.W. Chen, V. Greco, C.M. Ko, Bao-An Li, Phys. Rev. Lett. 90, 162701 (2003). 

\bibitem{ref82}
L.W. Chen, V. Greco, C.M. Ko, Bao-An Li, Phys. Rev. C 68, 014605 (2003). 

\bibitem{ref83}
L.W. Chen, C.M. Ko, Bao-An Li, Phys. Rev. C 69, 054606 (2004). 

\bibitem{ref84}
R. Ghetti et al., Phys. Rev. C 69, 03160 (2004). 

\bibitem{ref85}
R. Ghetti and J. Helgesson, Nucl. Phys. A 752, 480c (2005). 

\bibitem{ref86}
R. Ghetti et al., Phys. Rev. C 70, 034601 (2004).

\bibitem{ref87}
Y.D. Kim et al., Phys. Rev. C45, 387 (1992).

\bibitem{ref88}
R. Trockel et al., Phys. Rev. Lett. 59, 2844 (1987).

\bibitem{ref89}
Y. D. Kim et al., Phys. Rev. Lett. 67, 14 (1991).

\bibitem{ref90}
R. Bougault et al., Phys. Lett. B232, 291 (1994).

\bibitem{ref91}
D.R. Bowman et al., Phys. Rev. C52, 818 (1995).

\bibitem{ref92}
T. Glasmacher et al., Phys. Rev. C50, 952 (1994).

\bibitem{ref93}
R. Popescu et al., Phys. Rev. C58, 270 (1998).

\bibitem{ref94}
L. Beaulieu et al., Phys. Rev. Lett. 84, 5791 (2000).

\bibitem{ref95}
Ph. Chomaz et al., Phys. Rev. Lett. 73, 3512 (1994).

\bibitem{ref96}
M. Parlog et al., Eur. Phys. J. A 25, 223 (2005).

\bibitem{ref97}
E. Bauge et al., Phys. Rev. Lett. 70, 3705 (1993).

\bibitem{ref99}
E. Elmaani et al., Nucl. Instr. and Meth. A313, 401 (1992).

\bibitem{ref98}
E. Cornell et al., Phys. Rev. Lett. 75, 1475 (1995). 

\bibitem{ref100}
W.A. Friedman, Phys. Rev. C42, 667 (1990).

\bibitem{ref101}
E. Cornell et al., Phys. Rev. Lett. 77, 4508 (1995). 

\bibitem{ref102}
G. Tabacaru et al., Nucl. Phys. A764, 371 (2006).

\bibitem{ref103}
M. Colonna et al., Phys. Rev. C51, 2671 (1995).

\bibitem{ref104}
A. Guarnera et al., Phys. Lett. B 403, 191 (1997).

\bibitem{ref105}
N. Marie et al. (INDRA collaboration), Phys. Rev. C58, 256 (1998).

\bibitem{ref106}
P. Staszel et al., Phys. Rev. C63, 064610 (2001).

\bibitem{ref107}
S. Hudan et al., 
in \textit{Proceedings of the $XXXVIII^{th}$ 
International Winter Meeting on Nuclear Physics}, Bormio, Italy, 2000,
ed. by I.~Iori, A.~Moroni, Ricerca Scientifica ed Educazione 
Permanente Suppl. \# 116 (Milano 2000), p. 443.

\bibitem{ref108}
S. Hudan, Ph.D. thesis, Universit{\'e} de Caen, (2001) GANIL T01 07.

\bibitem{ref109}
S. Hudan et al. (INDRA collaboration), Phys. Rev. C67, 064613 (2003).

\bibitem{ref110}
O. Tirel, Ph.D. thesis, Universit{\'e} de Caen, (1998), GANIL T 98 02.

\bibitem{ref111}
R. Nebauer and J. Aichelin, Nucl. Phys. A650, 65 (1999).

\bibitem{ref112}
R. Nebauer et al., (INDRA Collaboration) Nucl. Phys. A658, 67 (1999).

\bibitem{ref113}
D.H.E. Gross, Rep. Prog. Phys. 53, 605 (1990).

\bibitem{ref114}
A. Ono et al., Phys. Rev. Lett. 68, 2898 (1992).

\bibitem{ref115}
A. Ono et al., Prog. Theor. Phys. 87, 1185 (1992).

\bibitem{ref116}
A. Ono et al., Phys. Rev. C 59, 853 (1999).

\bibitem{ref117}
Ph. Chomaz et al.,  Phys. Lett. B 254, 340 (1991).

\bibitem{ref118}
A.S. Botvina et al., Nucl. Phys. A475, 663 (1987).

\bibitem{ref119}
J.P. Bondorf et al., Phys. Rep. 257, 133 (1995).

\bibitem{ref120}
A.H. Raduta and A.R. Raduta, Phys. Rev. C55, 1344 (1997). 

\bibitem{ref121}
A.H. Raduta and A.R. Raduta, Phys. Rev. C56, 2059 (1997).

\bibitem{ref122}
A.H. Raduta and A.R. Raduta, Phys. Rev. C59, 323 (1999).

\bibitem{ref123}
D. Cussol, Phys. Rev. C68, 014602 (2003).

\bibitem{ref124}
A. Strachan and C. Dorso, Phys. ReV. C59, 285 (1999).

\bibitem{ref125}
C. Dorso and J. Randrup, Phys. Lett. B 301, 328 (1993).

\bibitem{ref126}
C. Escano et al., Proc. IWM2003 on Int. Work. On Multifragmentation and related topics, p. 197 (2003).

\bibitem{ref127}
D.W. Stracener et al., Nucl. Instrum. Meth. In Phys. Res. A 294, 485 (1990).

\bibitem{ref128}
A. Bonasera et al., Phys. Rep. 243, 1 (1994).


\bibitem{ref130}
S. Levit and P. Bonche, Nucl. Phys. A437, 426 (1985).

\bibitem{ref131}
S.E. Koonin and J. Randrup, Nucl. Phys. A474, 173 (1987).

\bibitem{ref132}
R.J. Charity et al., Nucl. Phys. A483, 371 (1988).

\bibitem{ref133}
T. Nayak et al., Phys. Rev. C45, 132 (1992).

\bibitem{ref134}
O. Ono, S. Hudan, A. Chbihi and J.D. Frankland, Phys. Rev. C66, 014603 (2002).

\end{thebibliography}

\end{document}